\documentclass[acmsmall,nonacm,screen]{acmart}

\AtBeginDocument{%
  \providecommand\BibTeX{{%
    \normalfont B\kern-0.5em{\scshape i\kern-0.25em b}\kern-0.8em\TeX}}}

\setcopyright{none}

\usepackage{wrapfig}
\usepackage{csquotes}

\usepackage{amsthm}
\usepackage{subcaption} 
\usepackage{booktabs}
\usepackage{multirow}
\usepackage{comment}
\usepackage{tablefootnote}

\theoremstyle{definition}
\newtheorem{definition}{Definition}[section]

\let\vec\mathbf
\usepackage{xcolor}

\begin{document}

\title{User-Specific Bicluster-based Collaborative Filtering: Handling Preference Locality, Sparsity and Subjectivity}
\let\vec\mathbf

\author{Miguel G. Silva}
\affiliation{%
  \institution{LASIGE and Faculdade de Ciências, Universidade de Lisboa}
  \streetaddress{Campo Grande 016, 1749-016}
  \city{Lisbon}
  \country{Portugal}}
\email{mmgsilva@ciencias.ulisboa.pt}

\author{Rui Henriques}
\affiliation{%
  \institution{INESC-ID and Instituto Superior Técnico, Universidade de Lisboa}
  \streetaddress{Av. Rovisco Pais 1, 1900-001}
  \city{Lisbon}
  \country{Portugal}}
\email{rmch@tecnico.ulisboa.pt}

\author{Sara C. Madeira}
\affiliation{%
  \institution{LASIGE and Faculdade de Ciências, Universidade de Lisboa}
  \streetaddress{Campo Grande 016, 1749-016}
  \city{Lisbon}
  \country{Portugal}}
\email{sacmadeira@ciencias.ulisboa.pt}

\renewcommand{\shortauthors}{Silva and Henriques, et al.}

\begin{abstract}
Collaborative Filtering (CF), the most common approach to build Recommender Systems, became pervasive in our daily lives as consumers of products and services. However, challenges limit the effectiveness of Collaborative Filtering approaches when dealing with recommendation data, mainly due to the diversity and locality of user preferences, structural sparsity of user-item ratings, subjectivity of rating scales, and increasingly high item dimensionality and user bases. To answer some of these challenges, some authors proposed successful approaches combining CF with Biclustering techniques.

This work assesses the effectiveness of Biclustering approaches for CF, comparing the impact of algorithmic choices, and identifies principles for superior Biclustering-based CF. As a result, we propose USBFC, a Biclustering-based CF approach that creates user-specific models from strongly coherent and statistically significant rating patterns, corresponding to subspaces of shared preferences across users. Evaluation on real-world data reveals that USBCF achieves competitive predictive accuracy against state-of-the-art CF methods. Moreover, USBFC successfully suppresses the main shortcomings of the previously proposed state-of-the-art biclustering-based CF by increasing coverage, and coclustering-based CF by strengthening subspace homogeneity. 
\end{abstract}



\maketitle

\section{Introduction}
\label{introduction}

Decision making plays a vital role in everyone's lives. 
As an attempt to cope with massive range of options, there has been large academic and industry interest in automatically recommending items to individuals since last century. Spotify, Amazon, Netflix, and Facebook are some popular platforms that actively use recommender systems \citep{CF-Survey-Ekstrand-2011}. From e-commerce to online advertisement, these systems are unavoidable in our daily online journeys to suggest items in a personalized way. Collaborative Filtering (CF) approaches, firstly proposed by \cite{CF-UserBasedCF-Goldberg-1992}, are currently seen as the widest implemented and most mature of the technologies to build recommender systems. Given a set of observed item ratings, CF aims at estimating unknown preferences based on the assumption that users with similar preferences in the past will yield similar preferences in the future. Despite the role of Collaborative Filtering, significant challenges limit its effectiveness, including the diversity and locality of user preferences, the structural sparsity of user-item ratings, the subjectivity of rating scales, and the increasingly large user and item bases \citep{CF-Survey-Ekstrand-2011,Survey-CF-Su-2009}. 

To address the diversity of user profiles, reduce the dimensionality and minimize rating sparsity, 
matrix factorization and clustering approaches have been combined within CF approaches for two decades \citep{CF-Survey-Ekstrand-2011}. 
However, traditional clustering techniques are typically applied to either group users or items separately. 
In real-world CF scenarios, the preferences of a subset of users is frequently only significantly correlated on a subset of the overall items, and vice versa \citep{BC-Survey-Kelvin-2013}. \textit{Biclustering} performs clustering in two dimensions simultaneously, being able to find these local preference patterns that correspond to data subspaces (biclusters) \citep{BC-Survey-Sara-2004}. Although biclustering has been originally proposed in biomedical domains \citep{BC-geneexpression-Church-2000, BC-geneexpression-Sara-2010, BC-geneexpression-Gupta-2010}, 
it increasingly shows promising results in the recommendation domain \citep{BiclustCF-scalablecf-george-2005,BCF-impactbiclusteringcf-Singh-2018, BCF-improvetopnrec-Feng-2020}. 
In this context, \cite{BiclustCF-scalablecf-george-2005} introduced co-clustering as a tool to scale Collaborative Filtering. Despite efficient, their approach can incur in predictive errors associated with the difficulty in guaranteeing a strong homogeneity in the found sub-spaces (co-clusters). More recently, \cite{BCF-impactbiclusteringcf-Singh-2018} proposed a novel biclustering-based CF approach that outperformed state-of-the-art rating prediction approaches, however, yielding a small coverage of unknown rating estimates. 

To address these challenges, this work proposes 
a novel CF approach, referred as \textit{User-specific Bicluster-based Collaborative Filtering} (USBCF), that identifies subspaces of shared preferences to create user-specific small and denser user-item matrices, which are then used to guide the training of traditional CF models. USBCF yields three major contributions of interest:
\begin{itemize}
    \item[--] state of the art biclustering searches for the discovery of high-coverage solutions of rating patterns with 
    statistical significance guarantees; 
    \item[--] superior matching of user preferences against the found rating subspaces by taking into account both the matching extent and fitness of preferences;
    \item[--] higher predictive coverage without hampering predictive accuracy by placing primary rating estimates from flexible biclustering structures, followed by secondary rating estimates from checkboard coclustering structures.
\end{itemize}

The proposed approach is assessed against baseline CF methods and 
peer biclustering-based CF approaches using reference real-world rating data. 
The gathered results show that the proposed approach successfully surpasses the limited coverage of state-of-the-art biclustering-based approaches 
and improves the rating prediction accuracy of traditional memory-based approaches.

The manuscript is organized as follows. Section 2 provides essential background on CF and Biclustering. Section 3 surveys relevant work on biclustering-based CF. Section 4 proposes USBCF, introducing the principles to address current shortcomings. Section 5 assesses USBCF against state-of-the-art peer approaches in real data, discussing its behavior. Finally, concluding remarks and future directions are drawn.

\section{Background}

\subsection{Collaborative Filtering}

\textit{Collaborative Filtering} (CF) aims at recommending items to users from available user-item rating data. Direct ratings are submitted by users explicitly, while indirect ratings can be inferred from their interactions with the system. User ratings can be further classified according to its domain -- unary (positive only), binary, ordinal or numerical \citep{CF-ExploitTimeCF-VinagreJoaoJorge-2015}. The set of ratings forms a sparse matrix, often referred to as U-I rating interaction matrix, with its sparsity being defined as the prevalence of unknown ratings.

\begin{definition}
Considering a set of $n$ users $\mathcal{U}=\{u_1,... , u_n\}$, a collection of $m$ items $\mathcal{I}=\{i_1,... ,i_m\}$, \textbf{user-item rating data} $\mathcal{D}$ is a set of preferences $r_{ui}$, each corresponding to a unary, binary, ordinal or numerical rating placed by user $u$ on item $i$. 

Given $\mathcal{D}$, \textbf{collaborative filtering} (CF) aims at estimating unknown preferences, $\hat{r}_{ui}$. Collaborative filtering can be reduced to specific users or items of interest, referred as active users or active items.
\end{definition}

CF gives rise to \textit{Predictive} and \textit{Recommendation} stances. In the Predictive stance, unknown ratings are estimated for a given user and item based on available ratings \citep{CF-EvaluateCF-Herlocker-2004}. As for the \textit{recommendation} tasks, the system usually generates and provides an ordered list of \textit{n} items, known as Top-\textit{N} recommendation list, which is a list including the most relevant/useful items for the user. This subsequent Recommendation task, often involves dissimilarity criteria among the returned items or between the known and top items \citep{CF-EvaluateCF-Herlocker-2004}.

Collaborative Filtering methods are usually divided into two main classes, the \textit{memory-based} and the \textit{model-based} \citep{Survey-CF-Su-2009}. The memory-based algorithms use the user-item system ratings directly to predict ratings for new items. 
In this context, user-based (item-based) systems predict the rating of an item by a particular user using a simple estimator over the set of ratings of similar users (items). 
Some challenges limit memory-based CF effectiveness when dealing with recommendation data, mainly due to the vast amounts of data and their sparse nature. Model-based approaches try to overcome the limitations of memory-based CF by learning predictive models from the available data and later use them to predict users' ratings for new items.

\subsection{Clustering}

\textit{Clustering} has been largely considered in Collaborative Filtering \citep{CF-Survey-Ekstrand-2011}. In the context of \textit{memory-based CF}, clustering has been applied to precompute groups of users (items) in user-based (item-based) approaches, promoting interpretability and testing time efficiency by assessing the similarity of a new (user) against the centroids of the precomputed clusters. Alternatively, clustering has been largely applied to \textit{model-based CF}, where the task of learning a predictive model from the complete rating data space is reduced to each cluster of users (items) produced from clustering, facilitating the learning task in accordance with group-wise preferences. 


Irrespective of the approach, clustering relies on similarity functions to estimate proximity between users (items). Popular examples robust to missing rates include cosine vector similarity and Pearson correlation \citep{CF-Survey-Ekstrand-2011}. 

Despite the relevance of the clustering-based CF, clustering mislay valuable information because they can only be applied either to the user or item dimension separately, disregarding the rich structure of the other dimension and thus preventing the identification of shared local preferences.

\subsection{Biclustering}

\begin{figure*}[h]
    \centering
    \includegraphics[width=0.95\linewidth]{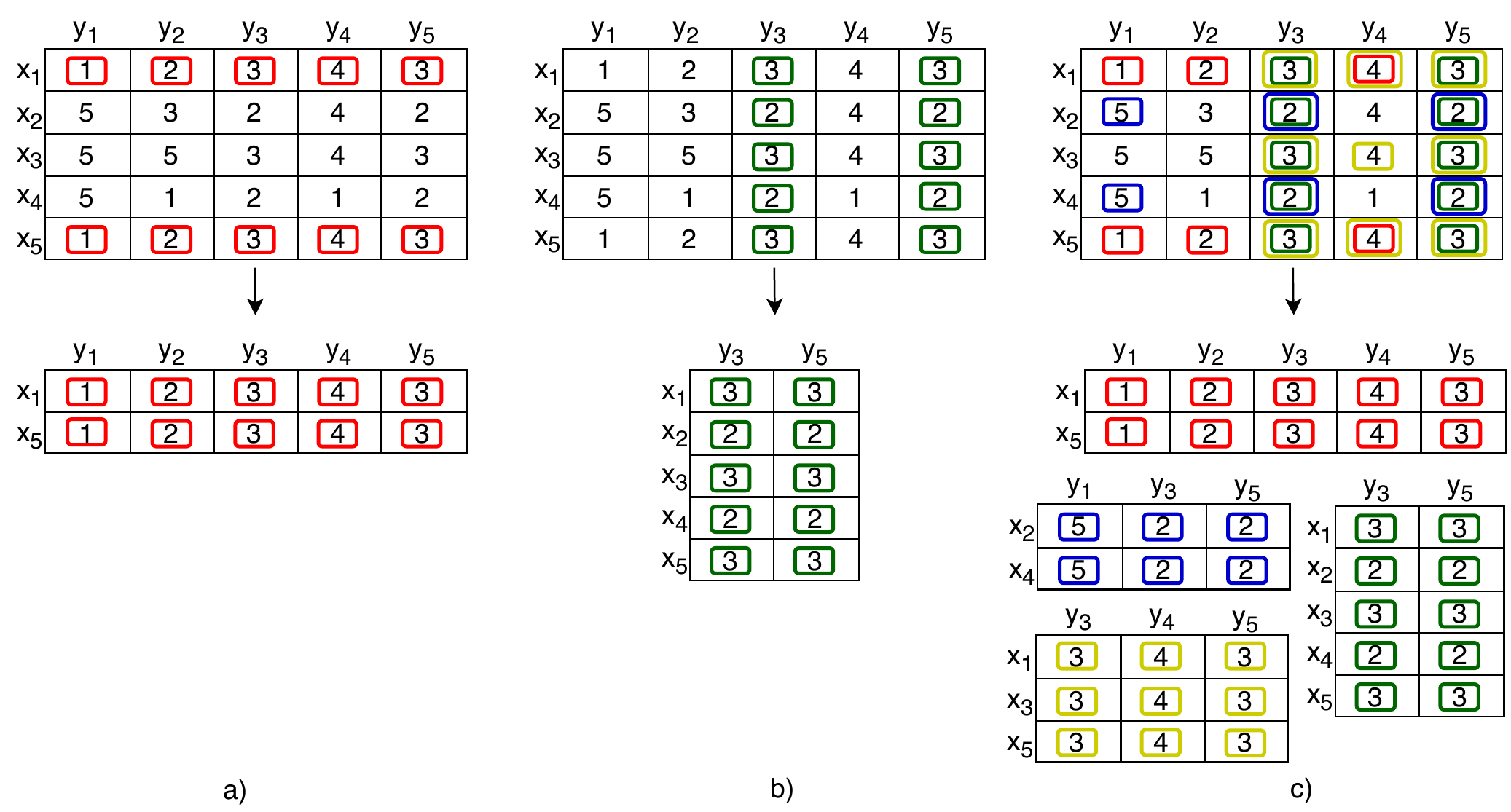}
    \caption{\small \mbox{Clustering (a-b) vs. biclustering (c) solutions over ordinal user-item ratings.}}
    \label{fig:biclusteringvsclustering}
\end{figure*}

In real-world applications with large item collections, shared preferences between a group of users (items) are only correlated on a subset of items (users). In this context, computing similarity against the whole dimension neglects the diversity of user preferences along item classes and therefore misses groups of users with correlated preferences on specific subsets of items. Opposed to one-way clustering, biclustering is thus a technique able to cluster users and items simultaneously, producing subspaces (biclusters) with coherent preferences as illustrated in Fig.\ref{fig:biclusteringvsclustering}. 

\begin{definition}
Given user-item rating data $\mathcal{D}$, a \textbf{bicluster} $B=(U,I)$ is a subspace given by a subset of users, $U\subseteq \mathcal{U}$, and a subset of items, $I\subseteq \mathcal{I}$. The \textbf{biclustering} task aims to find a set of biclusters $\mathcal{B}=\{B_1,..,B_q\}$, such that each bicluster $B_k$ satisfies specific criteria of homogeneity \citep{mypr}, dissimilarity and statistical significance \citep{Henriques2017}. 
\end{definition}

Biclusters capture local rating patterns corresponding to correlated preferences for a subset of users and items. The biclustering \textbf{homogeneity} criteria determines the \textit{structure}, \textit{coherence} and \textit{quality} of a biclustering solution. The \textbf{structure} is described by the number, size, shape and position of biclusters. Flexible structures of biclusters are characterized by an arbitrary number of (possibly overlapping) biclusters.  The \textbf{coherence} of a bicluster is given by the observed correlation of ratings (\textit{coherence assumption}) and, in the presence of numerical preferences, the allowed deviation from expectations (\textit{coherence strength}). Finally, the \textbf{quality} of a bicluster is defined by the type and amount of tolerated noise. 
\newpage 

\begin{definition}{}\label{def:bic_coherence}
Given a real-valued dataset, let the elements in a bicluster $r_{ui} \in (U,I)$ have \textbf{coherence} across observations given by $r_{ui}=k_i+\gamma_u + \eta_{ui}$, where $k_i$ is the expected rating for item $i$, $\gamma_u$ is the adjustment for user $u$, and $\eta_{ui}$ is the noise factor. A bicluster satisfying a specific coherence strength, $\delta \in \mathbb{R}^+$, has values described by $r_{ui}=k_i+\gamma_u + \eta_{ui}$ and $\eta_{ui} \in [-\delta/2,\delta/2]$.
\end{definition}

\begin{figure}[b]
\vskip -0.2cm
    \centering
    \includegraphics[width=0.77\linewidth]{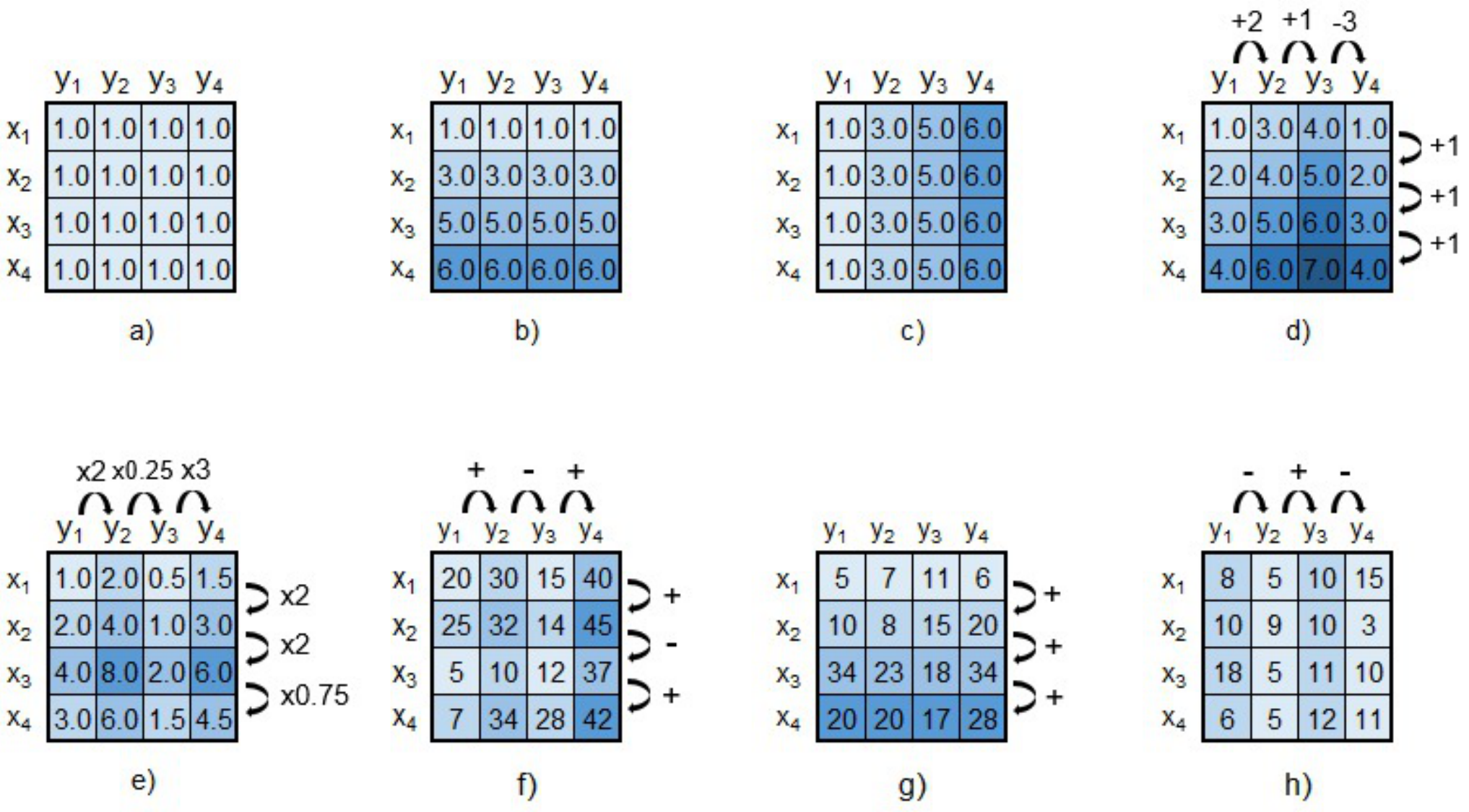}
    \caption{\footnotesize Illustrative forms of coherence \citep{BC-Survey-Sara-2004}: a) constant value, b) constant-values on rows, c) constant-values on columns, d) coherent values (additive model), e) coherent values (multiplicative model), f) overall coherent evolution, g) coherent evolution on the rows, h) coherent evolution on the columns.}
    \label{fig:biclusttypes}
\vskip -0.2cm
\end{figure}

\theoremstyle{definition}
\begin{definition}{}
Given a bicluster B=(U,I) with coherence in accordance with \autoref{def:bic_coherence}, the $\gamma_u$ factors define the coherence assumption: \textbf{constant} pattern on rows when $\gamma_u=0$ and \textbf{additive} pattern on rows when $\gamma_u \neq 0$. 

The bicluster \textbf{pattern} $\varphi_B$ is the set of expected values in the absence of adjustments and noise, $\varphi_B=\{k_i\mid i \in I\}$.
\label{patterndef}
\end{definition}

Commonly pursued forms of coherence are constant values across the subspace, users or items. When considering forms of numerical and ordinal feedback, biclusters can further accommodate additive, multiplicative or order-preserving factors. Fig.\ref{fig:biclusttypes} illustrates these different types of biclusters.

A bicluster is \textbf{statistically significant} if its probability to occur deviates from expectations (i.e. is unexpectedly low against a null data model). 

Let $\mathcal{B}$ be the set of biclusters that satisfy a given homogeneity and statistical significance criteria, $(U,I)\in\mathcal{B}$ is a \textbf{maximal bicluster} iff there is no other bicluster $(U',I')$ such that $U\subseteq U' \wedge I\subseteq I'$ satisfying the given criteria. Although an \emph{optimal biclustering solution} is one containing all maximal biclusters, the biclustering formulation can further include \textbf{dissimilarity} criteria to reduce the size of outputs and minimize redundancies. 

Even when considering unary rating data, biclustering is a well-established NP-hard task. As a result, many biclustering algorithms use heuristic mechanisms (producing sub-optimal solutions) \citep{BC-Survey-Sara-2004}, and place restrictions on the allowed homogeneity criteria.

\subsection{Contrasting biclustering with coclustering and dimensionality reduction}

The biclustering task is not equivalent to the \textbf{coclustering} task, also termed \textit{block clustering}. Coclustering is a restrictive form of biclustering requiring that all users and items belong to a subspace (exhaustive condition) and to a single subspace only (exclusive condition), producing a checkboard structure as a result. Although coclustering limits the inherent flexibility of the biclustering task \citep{kluger2003spectral}, it guarantees that all user-item pairs are included in a single subspace.

\begin{definition}
Given user-item rating data $\mathcal{D}$, \textbf{coclustering} aims to partition users and items, ($\mathcal{U}=\{U_1,..,U_r\},\mathcal{I}$=$\{I_1,..,I_s\}$), so that subspaces resulting from the intersecting partitions, $\mathcal{U}\times\mathcal{I}$ optimize some homogeneity criteria. 
\end{definition}

Dimensionality reduction of rating data became popularized after playing a significant role in the Netflix Prize competition \citep{CF-Dataset-Netflix}. Dimensionality reduction procedures are now pervasive in CF \citep{CF-Survey-Ekstrand-2011} as they offer a means of diminishing the sparsity of user-item rating data by placing data projections into denser spaces, and tackling generalization difficulties of model-based CF approaches caused by the arbitrarily-high dimensionality of item collections. 
Note, however, that most common forms of dimensionality reduction in CF are based on matrix factorization approaches, whose behavior resembles a restricted variant of coclustering approaches. In this context, they do not yield the inherent flexibility of biclustering stances in reducing the original space into an arbitrarily-high number of statistically significant subspaces capturing coherent preferences. 

\section{Related Work}
\label{relatedwork}

This section covers the most notorious contributions in the field of Collaborative Filtering incorporating subspace clustering techniques; from Predictive (section 3.1) to Recommendation (section 3.2) stances. 

\subsection{Rating prediction approaches}

In 2005, George and Merugu first introduced co-clustering as a tool to improve Collaborative Filtering  \citep{BiclustCF-scalablecf-george-2005}. In this work, they used weight Bregman co-clustering \citep{BC-Bregmancoclust-2004} to group users and items along a checkboard subspace structure, as firstly introduced by \cite{BC-Hartigan-partitionbiclust-1972}. Their co-clustering CF approach is based on the idea that the input data's missing ratings can be predicted using a suitable low parameter approximation of the input rating matrix, similar to CF approaches that rely on dimensionality reduction procedures such as singular value decomposition (SVD). 
The Bregman co-clustering returns non-overlapping subspace clusters covering the whole rating data space, as well as summary statistics derived from the co-clustering solution, to construct a matrix approximation for the input data matrix. For this particular application, the summary statistics are the rating averages of the users, items, and co-clusters. According to the authors, this permits a more reasonable rating approximation than just considering the average value of the co-cluster corresponding to the value we want to predict, which is plausible since it takes into account the biases of individual users and items. After computing the reconstructed approximate matrix, its values are considered for the rating prediction task. The authors designed incremental and parallel versions of the original co-clustering algorithm to build an efficient real-time CF mechanism, capable of updating the co-clusters as new users and ratings enter the system. The results of the work show that their approach achieves satisfactory accuracy compared with baseline matrix factorization models but at a lower computational cost.

In 2007, \cite{BCF-applybiclustcf-Castro-2007} suggested a biclustering CF methodology, using a heuristic immune-inspired biclustering technique, denoted BIC-aiNet, that finds biclusters with coherent values. Their methodology is based on two main stages: the biclusters' generation, and the similarity computation between the users and the resulting biclusters. After their artificial immune-inspired network generates the biclusters, they are used to predict how the active user would rate a specific item. The rating prediction is performed by searching for the biclusters that include the active user and item, and then, calculating the residue of each bicluster through mean-squared residue (MSR) \citep{BC-ChengChurch-2000}, 
    
\begin{equation}\label{eq:relatedwork-castro-residuemeasure}
    \mathrm{MSR(B)} = \frac{1}{|U||I|}\sum_{u\in U}\sum_{i\in I}(r_{ui}-r_{Ui}-r_{uI}-r_{UI})^2,
\end{equation}
\vskip 0.1cm
    
\noindent where $|U|$ is the number of users in the bicluster, $|I|$ is the number of items in the bicluster, $r_{ui}$ is the rating of user $u$ assigned to item $i$, $r_{Ui}$ and $r_{uI}$ represent the mean value of user $u$ and item $i$, respectively, and $r_{UI}$ is the mean value of the entire bicluster. Finally, the bicluster with the smaller residue value is selected, and the average of its movies' ratings is used as the prediction. This methodology was evaluated in both rating prediction and Top-\textit{N} recommendation scenarios, achieving better scores than the methods used for comparison, being the constant version of BCF \citep{BCF-nearestbicsconst-Symeonidis-2006}, one of those.

\cite{BCF-franca-2009} also used an immune inspired biclustering algorithm, MOM-aiNET \citep{BC-MOMainet-Coelho2008} to obtain biclusters used to predict missing ratings for recommendation purposes. Their algorithm generates biclusters with a controlled percentage of missing values. Then, they use a quadratic programming approach to predict those values by trying to minimize the mean-squared residue of the bicluster. This approach is viewed as an improvement of the one developed by \cite{BCF-applybiclustcf-Castro-2007} (described above), as the latter simply replaces the missing values in a bicluster by the average of the bicluster's values, not reflecting the trends obtained by coherent biclusters. Whereas, in this approach, the missing values are viewed as variables that should be obtained to minimize the residue.

\begin{figure}
        \centering
        \includegraphics[width=\linewidth]{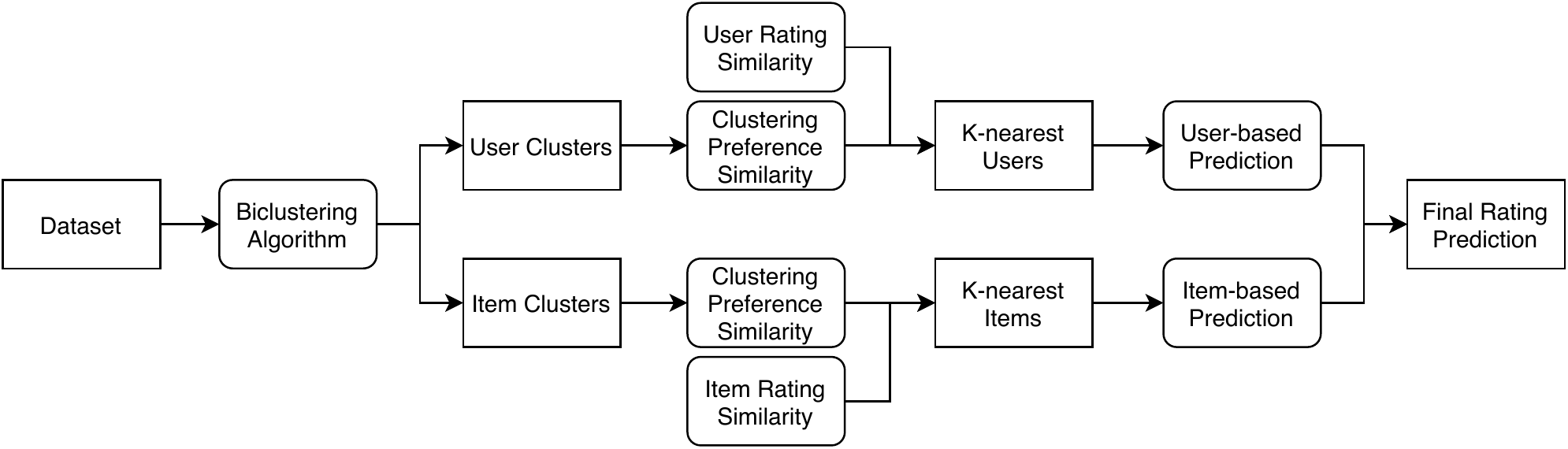}
        \caption{Outline of ITCCCF rating prediction approach by \cite{BCF-LiangLeng-2014}}
        \label{fig:CFbasedITCC-Liang2014}
    \end{figure}

A more recent work in the direction of biclustering CF was published in 2014 by \cite{BCF-LiangLeng-2014}, who proposed a CF algorithm utilizing information-theoretic co-clustering (ITCC) \citep{BC-ITCC-Dhillon-2003}. Their approach (ITCCCF) consists in performing two-way clustering to discover a set of user clusters and a set of item clusters. After that, they compute the preference of a user $u$ for the cocluster containing an item $i$, $B=(U,I)$, ($P_{u,J}$), for each pair user-item per cluster,  
    
    \begin{equation}\label{eq:userpref}
        P_{u,i} = \frac{|I_u \cap I_i|}{|I_u|},
    \end{equation}    
    
\noindent where $I_u$ are the items that the user $u$ rated and $I_i$ are the items from the cocluster containing item $i$. A matrix of user-cluster preferences is constructed and its values are used to obtain the cosine similarity between each pair of users, $sim_p(u_1,u_2)$. This approach then combines the previous clustering preference similarity with a rating similarity determined through Person's correlation coefficient to find the $k$ most similar users. The $k$-nearest users are selected and the User-based CF is used to generate the rating prediction. Moreover, the whole process is also repeated for the item clusters and the Item-based CF is used to generate a new rating prediction. The final rating prediction is a linear combination that fuses the ratings from the user-based and item-based approaches. In order to evaluate their approach, using real-world datasets, the authors compared it with 5 state-of-the-art methods, some already mentioned in this work (UBCF \citep{UserBasedCF-Resnick-1994}, IBCF \citep{ItemBasedCF-Sarwar-2001}, CBCF \citep{ClusteringCF-Sarwar-2002}, BCC \citep{BiclustCF-scalablecf-george-2005}), being able to surpass all of them regarding prediction accuracy. Figure \ref{fig:CFbasedITCC-Liang2014} provides an overview of how this fusion rating technique works.

In 2017, \cite{BCF-biclustcffusion-kantmahara-2017} also proposed a fusion-based approach called Nearest Biclusters Collaborative Filtering with Fusion (NBCFu) to address rating prediction using biclustering. In this approach, the xMOTIFs algorithm \citep{BC-xMotifs-Murali-2003} is used to generate the biclusters that are viewed as quality neighbors of users and items. After the biclusters' generation, the authors use the CjacMD measure \citep{BCF-biclustcffusion-kantmahara-2017-CjacMD-SIMMEASURE} (\autoref{eq:simmeasure-biclustcffusion-kantmahara-2017-CjacMD})  to obtain the K-nearest biclusters of the users and items. Considering a subspace of users $U\subseteq \mathcal{U}$, this similarity measure combines Mean Measure of Divergence (MMD) \citep{BCF-biclustcffusion-kantmahara-2017-MMD-SIMMEASURE}, Jaccard similarity, and cosine similarity,

    \vskip -0.2cm
    \begin{equation}\label{eq:simmeasure-biclustcffusion-kantmahara-2017-CjacMD}
        sim(u,U)_{JacMD}=sim(u,U)_{cos} + sim(u,U)_{Jaccard} + sim(u,U)_{MMD}.
    \end{equation}
\vskip 0.1cm
    
\begin{figure}
        \centering
        \includegraphics[width=\linewidth]{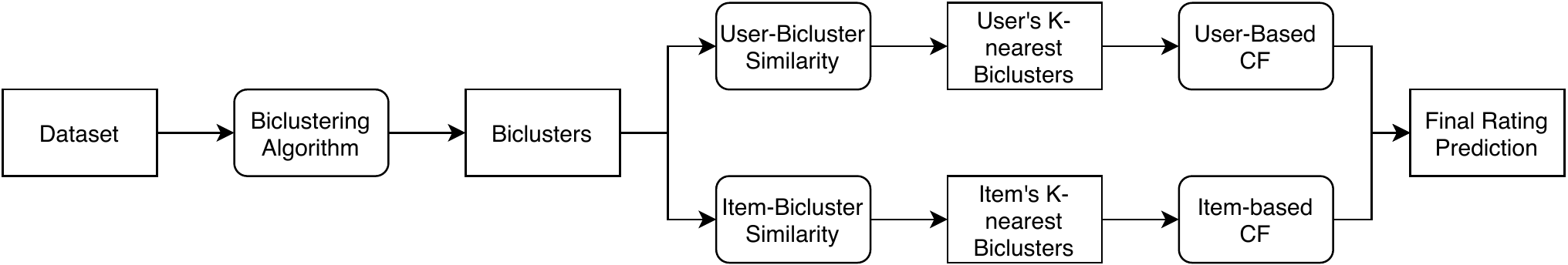}
        \caption{Outline of NBCFu rating prediction approach by \cite{BCF-biclustcffusion-kantmahara-2017}.}
        \label{fig:NBCFu-Kant2017}
    \end{figure}

After this step, User-based CF and Item-based CF using cosine similarity are applied to the users/items found in most similar biclusters. Finally, the rating prediction of both approaches is combined through a weighted sum, optimized using gradient descent, to generate a final and more accurate rating prediction. The authors claim that their approach has better accuracy results than some state-of-the-art CF methods, including SVD++ \citep{MatrixFactCF-svd++-koren2008}. \autoref{fig:NBCFu-Kant2017} shows an overview of NBCFu.

El-Nabarawy et al. \cite{BCF-artmap-elnabarawy-2016} examined the viability of biclustering ARTMAP (BARTMAP) \citep{BC-BARTMAP-XU-2011} for recommendation purposes. BARTMAP is a biclustering algorithm based on the Adaptive Resonance Theory neural network model \citep{BCF-art-elnabarawy-2016}. The ratings for an active user are predicted through a normalized weighted sum of the ratings that the remaining users gave to the item in the same bicluster, weighted by their correlation with the active user. The similarity score between an active user $u$ and another user $u'$, $sim(u,u')$, belonging to the same bicluster is given by the Pearson correlation coefficient of their ratings in the bicluster,

    \begin{equation}\label{eq:ratingpredartmapelnabarawy-2016}
        \hat{r}_{ui} = \bar{r}_{uI}+\frac{\sum\limits_{u' \in U }(r_{u'i}-\bar{r}_{u'I}) \cdot sim(u,u')}{\sum\limits_{u' \in U}|sim(u,u')|}.
    \end{equation}
\vskip 0.05cm
This formula can be seen an adaptation of the rating estimates used in User-based CF but restricted to a subspace. 
Using this prediction approach, the users with the highest positive correlation with the active user, have the most impact on the prediction. The algorithm's performance was compared against other collaborating filtering techniques, and it performed similarly to the previously mentioned approach, BIC-aiNet \citep{BCF-applybiclustcf-Castro-2007}.

Recently, \cite{BCF-impactbiclusteringcf-Singh-2018} introduced a new biclustering-based CF technique, BBCF. In this work, the authors use biclustering as a preprocessing step to scale CF approaches. Once the biclustering algorithm is executed, the users are compared with the found biclusters based on the items they have in common, 
    \vskip -0.2cm

    \begin{equation}\label{eq:similarityuserbiclusteruser}
        sim(u,B_k) = \frac{|I_u \cap I_k|}{|I_k|}.
    \end{equation}
    \vskip 0.1cm

Then, similarities between users and biclusters are weighted with the number of users in the bicluster so that biclusters with more users are privileged,
    
    \begin{equation}\label{eq:WF}
        \mathrm{WF}(u,B_k) = sim(u,B_k) \times |U_k|.
    \end{equation}
\vskip 0.08cm

After this step, the K-nearest biclusters per user are merged, creating a larger subspace referred as Nearest Neighbor Setup (NSS). The NSS of each user is viewed as a denser subspace of the U-I matrix, including a subset of users similar to the respective user along a subset of overall items. Finally, a classic CF approach, such as the Item-based, is applied over the personalized NSS of each user to predict the missing ratings. According to the authors, this approach tackles the scalability and sparsity problems that deteriorate the performance of memory-based CF methods, through the usage of the personalized NSSs instead of the entire U-I matrix. The authors evaluated BBCF using QUBIC and Item-based CF as the underlying biclustering and CF approaches, respectively, 
yielding better scores in most scenarios. However, they also highlight a clear drawback of their methodology. They point out that their approach cannot generate a prediction/recommendation for as many users/items as traditional methods. This handicap occurs because the NSS subspaces used within the CF algorithm may not fully cover the original user-item rating data space. 
    
     \begin{figure}
        \centering
        \includegraphics[width=0.8\linewidth]{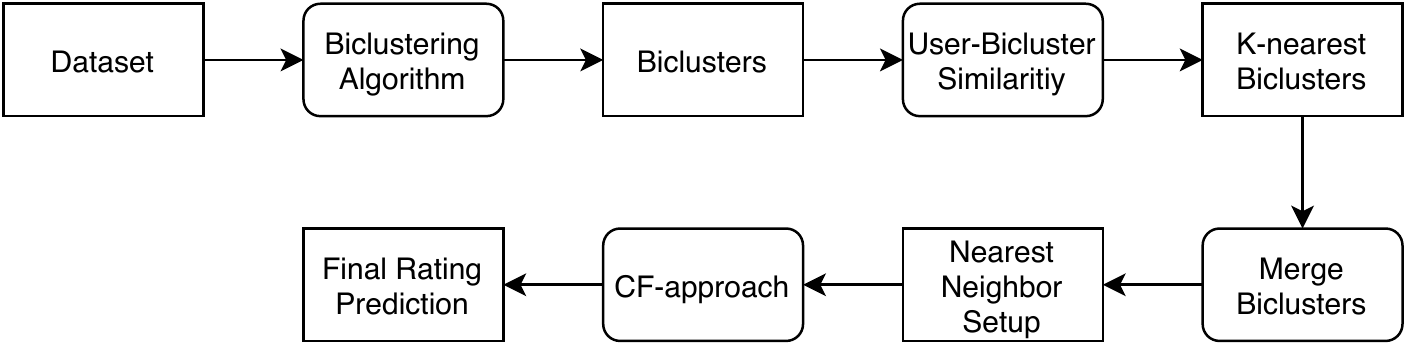}
        \caption{Outline of BBCF rating prediction approach proposed by \cite{BCF-impactbiclusteringcf-Singh-2018}.}
        \label{fig:BBCF-SinghMehotra-2017}
    \end{figure}
    
\subsection{Top-\textit{N} Recommendation Approaches}

\begin{figure}
        \centering
        \includegraphics[width=0.73\linewidth]{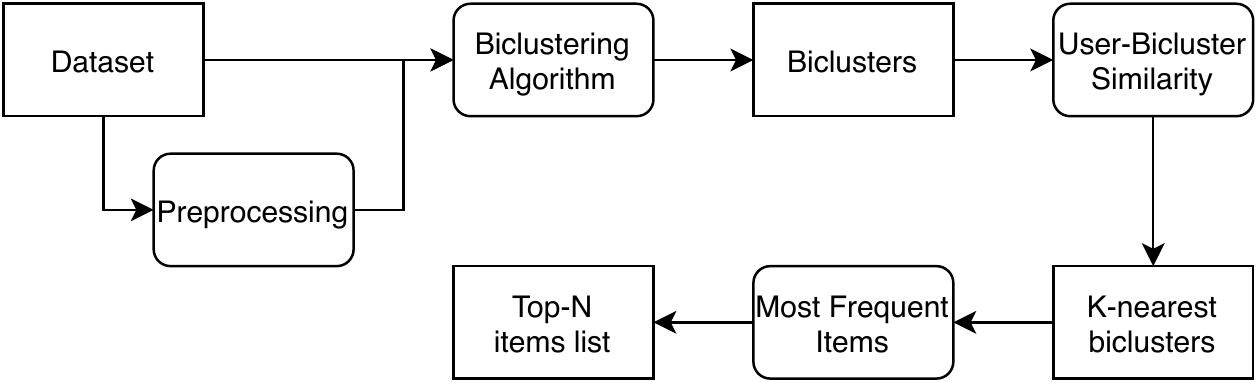}
        \caption{Outline of the NBCF top-\textit{N} recommendation approach by \cite{BCF-nearestbicsconstcoherent-Symeonidis-2008}.}
        \label{fig:CFNearestBicstopnrecommendation-symeonidis-2007}
    \end{figure}

\cite{BCF-nearestbicsconstcoherent-Symeonidis-2008} proposed a neighborhood-based CF algorithm (NBCF) that uses biclustering to improve scalability and accuracy of CF. They also created a similarity measure, defined in \autoref{eq:similarityuserbiclusteruser}, to identify biclusters that better reflect users' preferences. Their approach can be combined with biclustering algorithms that find subspaces with either constant or coherent values. The authors used Bimax and xMOTIFs, respectively.  Upon selecting biclusters with overall constant values, we can only discover sets of users and items with identical rating values. Whereas, a coherent pattern allows to find users and items correlated with more complex behaviors, such as users that, for a subset of items, exhibit coherent ratings. After the discovery of biclusters, they measure the similarity of the active user with each bicluster, finding the k-nearest-biclusters to the user. Then, to capture the influence that each item belonging to a bicluster has over the active user, they use Weighted Frequency (WF), as in \autoref{eq:WF}. Finally, Weighted Frequencies scores are summed per item. 
The final top-\textit{N} recommended items are the $N$ items with the highest sum of WF.  \autoref{fig:CFNearestBicstopnrecommendation-symeonidis-2007} summarizes the process of this top-\textit{N} recommendation approach. The authors reported that both proposed approaches -- constant and coherent biclustering assumptions -- achieved better results in terms of accuracy and efficiency than classic CF algorithms such as User-based CF \citep{UserBasedCF-Resnick-1994}, Item-based CF \citep{ItemBasedCF-Sarwar-2001}, and clustering-based CF (CBCF) \citep{ClustCF-xue-2005}. They also concluded that despite being slightly less efficient, the recommendation accuracy when considering coherent biclusters outperforms the baseline approach with constant biclusters.

Still in the context of top-\textit{N} recommendation tasks, \cite{BCF-biclustneighborhoodtopn-alqadah-2015} also  proposed a biclustering neighborhood-based CF method (BCN). In this work, the authors use properties from the field of Formal Concept Analysis (FCA) \citep{BC-biclustneighborhoodtopn-FCA-BOOK}. FCA is a specialized form biclustering aiming at discovering dense subspaces from rating data produced from unary or binary preferences \citep{BC-biclustneighborhoodtopn-FCA-PHDTHESIS-juniarta2019}. FCA algorithms can be adapted to enumerate and order subspaces, which can then be used to identify neighborhoods of closely related subspaces. Alqadah et al. took advantage of these FCA properties to build neighborhoods for active users. Their approach finds the \enquote{smallest} subspace containing the active user $u$, which is the subspace containing the fewest number of users and the greatest number of items.  Then, by exploring the subspace neighborhood through FCA properties, it finds similar users to $u$, and appends the items in the neighborhood to a candidate set of items. Finally, the candidate items are ranked by combining a global and a bicluster neighborhood similarity, and the top $N$ items are returned as recommendations. The global distance between a user $u$ and an item $i$, $g(u,i)$, measures the similarity between the user and the item in the entire matrix,
\vskip -0.05cm
\small
    \begin{equation}
        g(u,i)= \frac{\sum_{i_k \in \mathcal{L}}Jacc(U_i,U_k)}{|I'|},
    \end{equation}
    \normalsize
    \vskip 0.15cm

\noindent where $J(U_i,U_k)$ is the Jaccard index defined over the set of all users who interact with item $i$ and the set of those who interact with item $i_k$, and $\mathcal{I}'$ is the set of items belonging to all found subspaces. In contrast, the local/neighborhood similarity, $l(u,i)$, captures similarities between locally similar users and items, considering the bicluster similarity \citep{BC-biclustneighborhoodtopn-bicsimilarity} using subspaces in which $i$ occurs. The authors claim that their approach is superior to those computing the biclusters offline since, in BCN, they map a user to a bicluster on demand. However, the BCN framework was designed to work on implicit feedback recommendations, and thus it can only be applied to binary data. Nevertheless, to surpass this limitation, it could be an interesting research stream to study the potential of an FCA adaptation capable of dealing with numerical rating data. For instance, in a very recent thesis, \cite{BC-biclustneighborhoodtopn-FCA-PHDTHESIS-juniarta2019} proposed and FCA extension to deal with numerical matrices and discover more flexible types of biclusters which could be potentially relevant for recommendation purposes.


\section{User-Specific Biclustering-based Collaborative Filtering}
\label{sec:usbcf}

Grounded on the previously surveyed contributions and further empirical evidence, this section introduces a new approach, User-Specific Biclustering-based Collaborative Filtering (USBCF). 

The proposed USBCF approach is divided into four major steps. First, biclusters are mined from rating data using biclustering searches that guarantee the exhaustive discovery of flexible structures of statistically significant, and well-defined rating patterns (section 4.1). 
Second, and inspired by the Singh and Mehotra contributions \citep{BCF-impactbiclusteringcf-Singh-2018}, the found biclusters are used to create larger subspaces personalized for each user (section 4.2). These subspaces are denser and show stronger preference correlation than the original rating data. Third, we introduce a novel matching mechanism between an active user and the enlarged subspaces that considers both the coverage and correlation between the user and the biclusters (section 4.3). By integrating item coverage and correlation profile, USBCF guarantees that the selected subspaces per user are relevant, showing reduced propensity to contain divergent preferences. Fourth, and once the personalized subspace is created for a particular user, a well-established CF algorithm is used to originate a unique predictive model for each system user (section \ref{coclussec}). Finally, secondary rating estimates from checkboard coclustering structures are further collected to account for missing ratings that may fall outside each personalized subspace.

\begin{figure*}[h]
    \centering
    \includegraphics[width=\linewidth]{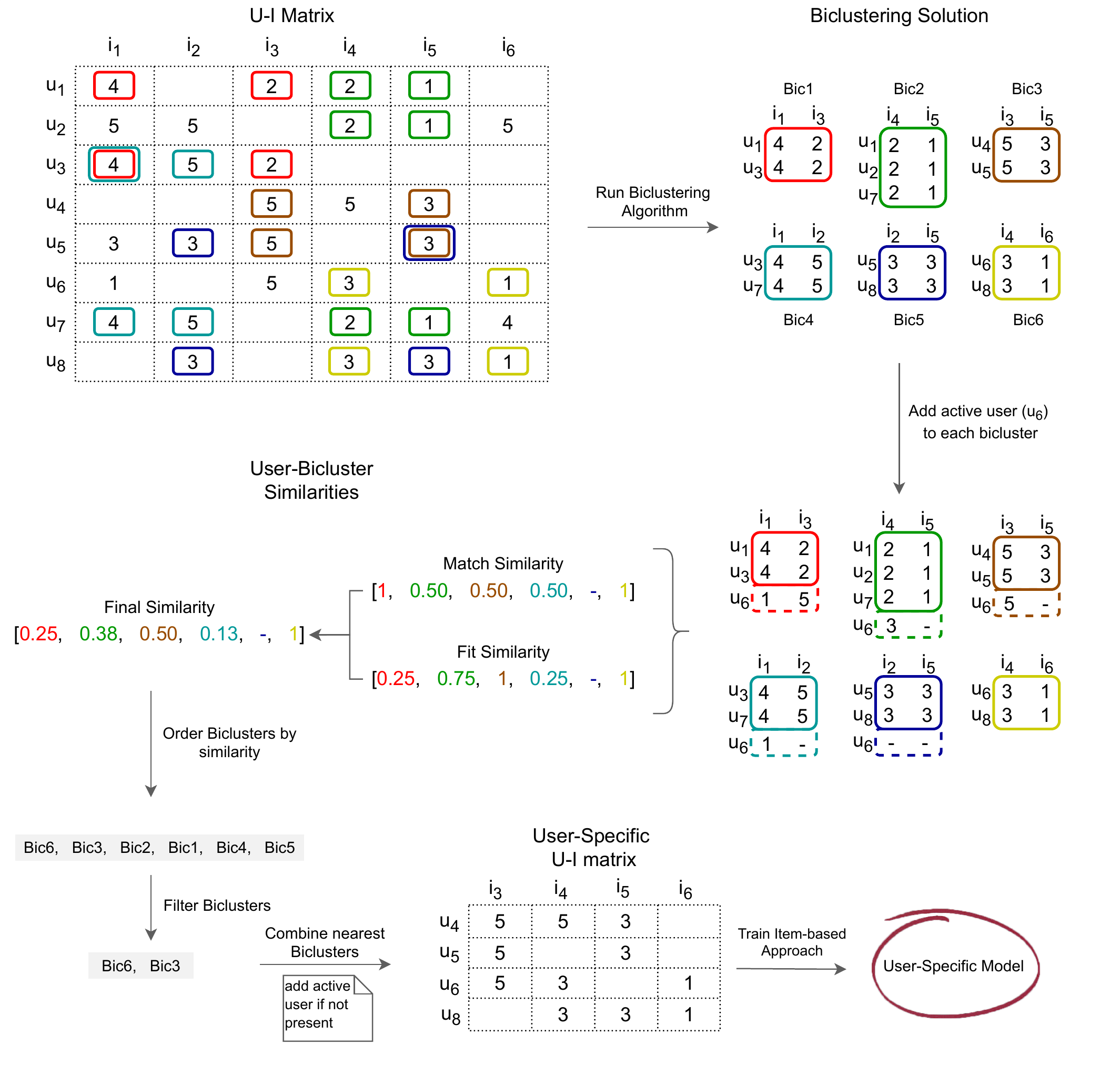}
    \caption{USBCF approach applied on a small U-I matrix.}
    \label{fig:usbcfapproach}
\end{figure*}

The methodology can be easily parallelized since the entire process in independent for each user. \autoref{fig:usbcfapproach} presents and overview of the USBCF methodology applied over an illustrative sample of user-item ratings and the corresponding user-specific CF model.

\subsection{Bicluster discovery in user-item rating data}

The first step of the approach consists of mining biclusters within user-item rating data, such that three major properties of interest are observed: i) nearly-exhaustive biclustering searches are employed to guarantee that all potentially relevant biclusters are identified, thus maximizing the coverage of the original rating dataset and, as a result, supporting the prediction of missing ratings; ii) the targeted type of biclusters (Fig.~2) should guarantee that meaningful and well-delineate preference patterns are retrieved, allowing coherent preference variations to take place on different items of a subspace; and iii) the retrieved biclusters should pass statistically significance tests, thus minimizing the susceptibility to use false positive preference patterns (retrieved preference patterns that occur by chance) in the prediction process. 

Existing biclustering-based CF approaches generally fail to satisfy some of the properties due to the underlying biclustering algorithmic choices. In this work, we select QUBIC2\footnote{\url{https://github.com/OSU-BMBL/QUBIC2}}  \citep{BC-QUBIC2-2019} as the biclustering algorithm. QUBIC2 is an efficient nearly-exhaustive deterministic biclustering algorithm that identifies statistically significant scaling-patterned biclusters with constant-values on the rows. 

When applied over an U-I Matrix, QUBIC2 identifies groups of users who have the same preference for the bicluster items. The tolerated noise per bicluster can be further parameterized in QUBIC2. Following empirical evidence, we restrict the biclustering search to discover non-noisy maximal biclusters as noise and preference divergences can be accommodated in the subsequent subspace enlarging step (section 4.2). The minimum size of the biclusters is controlled by a parameter \textit{minCols}. Different values of \textit{minCols} may lead to different biclustering solutions. Instead of setting the \textit{minCols} parameter to a fixed value, in this work, we created an enriched biclustering solution that aggregates the solutions from running QUBIC2 with different \textit{minCols} setups. A detailed description of the algorithm can be found in QUBIC's original works \citep{BC-QUIBIC-2009,BC-QUBIC2-2019}.

\subsection{Biclustering neighborhood formation}

The goal of this second step is to create, for each user, a new data space with meaningful and personalized information. To do so, we take advantage of the biclusters found by the biclustering algorithm in the previous module. After the biclusters are generated, we can expect that some of those biclusters represent a user's preferences better than others. From this perspective, we try to find a \enquote{neighborhood} of biclusters. In other words, a subset of overall found biclusters such that each bicluster in this subset, $B\in\mathcal{B}$, has a preference pattern, $\varphi_B$, correlated with the observed preferences for the active user. Once this neighborhood of biclusters is identified, we followed the principles proposed by \cite{BCF-impactbiclusteringcf-Singh-2018} to create a personalized user-specific dataset. 

\begin{figure}[H]
    \centering
    \includegraphics[width=0.6\linewidth]{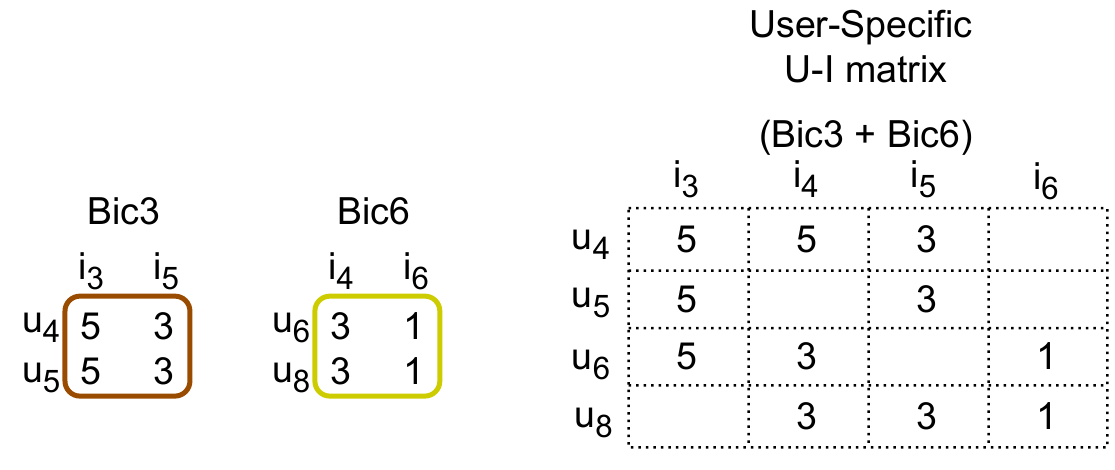}
    \caption{Example of aggregating two biclusters to create a new dataset.}
    \label{fig:usbcfmergebiclusters}
\end{figure}

The procedure to obtain the personalized dataset consists in merging all the users and items of the bicluster neighborhood, corresponding to a full outer join of the subspaces in the neighborhood. 
For example, considering $B_3=(U_3=\{u_4,u_5\},I_3=\{i_3,i_5\})$ and $B_6=(U_6=\{u_6,u_8\},I_6=\{i_4,i_6\})$ in \autoref{fig:usbcfapproach}, the union set of users and items of this neighborhood is $(U_N = \{u_1, u_2, u_6, u_7, u_8\}, I_N = \{i_4, i_5, i_6\}$. 
\autoref{fig:usbcfmergebiclusters} illustrates the process.

Finally, the active user is added to the personalized rating dataset when he is not amongst the set of users of his own space. This step is crucial to accommodate available preferences from the active user and to guarantee that the subsequent application of Collaborative Filtering approaches can generate recommendations for his omissive preferences.

\subsection{Similarity between individual and subspace preferences}

To identify neighborhoods for personalized data creation, the subset of biclusters with preference patterns correlated with the preferences of an active user need to be identified. \cite{BCF-nearestbicsconstcoherent-Symeonidis-2008} introduced the concept of similarity between a bicluster and a user. However, their similarity is solely based on the rated items' interception, thus being unable to guarantee that the preferences of the active user are correlated with the preferences of the users in the biclusters. Understandably, this is an undesirable condition in the presence of binary, ordinal or numeric rating data.  

To address this limitation, we propose a new user-bicluster similarity that, besides the rated items' interception, also considers the values of ratings. 
Following empirical evidence, this similarity is calculated through the product of two distinct scores which we refer to as \textbf{items' match} ($sim_{match}$) and \textbf{pattern fit} ($sim_{fit}$), respectively.
 
\begin{figure}[H]
    \centering
    \includegraphics[width=0.46\linewidth]{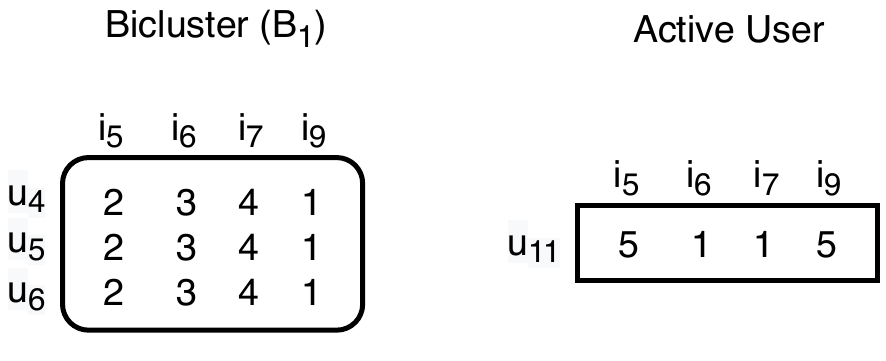}
    \caption{Example of a bicluster and an active user with distinct rating pattern.}
    \label{fig:bicactiveuserdifferent}
\end{figure}

The \textbf{items' match score} captures the portion of bicluster items which were rated by the user, 
 
\begin{equation}\label{eq:sim_match}
    \mathit{sim}_{\mathit{match}}(u,B_k) = \frac{|I_u \cap I_k|}{|I_k|},
\end{equation}
\vskip 0.1cm

\noindent where $I_u$ are the items rated by the user \textit{u} and $I_k$ are the items in bicluster $B_k$. Its values range between [0,1], with 1 representing the maximum similarity score between a user and bicluster. 
Understandably, a similarity entirely based on rated items' is insufficient as it disregards the values of the ratings. For instance, in \autoref{fig:bicactiveuserdifferent}, we have an example of bicluster and an active user that rating all the items in the bicluster. Using similarity criteria solely based on \autoref{eq:sim_match}, we would yield a maximum similarity. However, the preferences between the active user and the ones from the given bicluster are divergent, thus the inclusion of such bicluster in the neighborhood would hamper the subsequent CF task. 

Given this, we introduce the \textbf{pattern fit score}, which is responsible for measuring the resemblance between the user rating pattern and the bicluster preference pattern (Def.~2.4). Given a user $u$ and a bicluster $B_k=(U_k,I_k)$, 

\begin{equation}\label{eq:sim_fit}
    \mathit{sim}_{\mathit{fit}}(u,B_k) = 1 - \frac{\mathit{RMSE}(\varphi_{B_k} , \varphi_{(u,I_u\cap I_k)} )}{r_{\max}-r_{\min}},
\end{equation}
\vskip 0.1cm


\noindent where $\varphi_{B_k}$ is the bicluster pattern (Def.~\ref{patterndef}), $\varphi_{(u,I_u\cap I_k)}$ is the user preferences along the items $I_k$ in the bicluster, $r_{max}$ ($r_{min}$) correspond to the maximum (minimum) observed rating, and RMSE corresponds to the root mean squared error. 
In sum, the pattern fit scores corresponds to the normalized root-mean-squared differences between the preferences defined by the subspace pattern and the ratings of the user.

The similarity score defined in \autoref{eq:sim_fit} considers a rating error applied to biclusters with a constant coherent pattern. However, this similarity could be easily adapted to different coherence types, such as order-preserving, by adopting a Spearman's rank correlation coefficient or a Kendall rank correlation coefficient, instead of the root mean squared error, to take into account order preservation in the patterns.

The final similarity between a user and a biclusters is the simple product between the scores previously defined,

\vskip -0.1cm
\begin{equation}
    \mathit{sim}(u,B_k) = \mathit{sim}_{\mathit{match}} \times \mathit{sim}_{\mathit{fit}},
\end{equation}
\vskip 0.1cm

\noindent as revealed by empirical evidence from the experimental comparison of weighted sums, $\sum_i \alpha_i x_i$, and weighted products, $\prod_i x_i^{\alpha_i}$.

Both \textbf{items' match score} and \textbf{pattern fit score} values range between [0,1], so its product will also be in the same range. The product operation further establishes the relevance for both components to yield good levels for a bicluster to be seen as relevant for an active user. In other words, a poor value in one of the components will result in a low similarity score. 

A \textit{minSim} threshold is finally considered to filter the relevant biclusters for an active user. Only the biclusters with similarity with the user above the threshold are selected to produce its personalized dataset.


\subsection{Learning User-Specific Recommendation Models}
\label{coclussec}

Once the personalized datasets are created, a Collaborative Filtering algorithm is subsequently trained to create a unique recommendation model for each individual. Although the USBCF approach can be parameterized with any Collaborative Filtering approach, Item-based CF approaches are a particularly suitable option due to the inherent properties of the personalized data spaces. Despite their inherent simplicity, Item-based CF approaches are one of the form CF approaches and still recognized for producing state-of-the-art rating predictions. USBCF generates smaller and denser U-I matrices which, when allied with these approaches, allow the models to perform the recommendation tasks more effectively.

It is possible that the personalized data spaces do not cover the entire recommendation space. For instance, in \autoref{fig:usbcfapproach}, the personalized dataset does not contain items $i_1$ and $i_2$, originating a user-model with limited coverage capability, i.e. missing preferences for the items not included in the personalized data space cannot be estimated using the personalized. These situations occur when there are no shared local preferences containing those items, which poses a particularly difficult predictive setting irrespective of the selected CF approach. 

To address this challenge, it is possible to find subspaces with lower homogeneity, specifically an exhaustive partitioning of the original space into a set of subspaces that cover the entire itemset. Following the principles introduced by \cite{BiclustCF-scalablecf-george-2005}, USBCF relies on the previously proposed coclustering approach to estimate the typical small fraction of items falling into this condition, resorting to its default behavior for items included in the personalized data spaces.  




\section{Experiments and Results}

This section experimentally assesses the proposed 
approach against state-of-the-art peers. In particular, we aim to answer the following research questions:

\begin{itemize}

    \item \textbf{RQ1.} How does USBCF compare against baseline CF approaches in terms of quality of predictions and coverage capability?
    \item \textbf{RQ2.} How does the biclustering search and minimum user-bicluster similarity threshold impact USBCF performance?
    \item \textbf{RQ3.} Can USBCF surpass the main limitations presented by the state of the art Biclustering-based CF (BBCF) \citep{BCF-impactbiclusteringcf-Singh-2018}?
    
\end{itemize}

\subsection{Experimental Setting}

\subsubsection{Experimental Platform and Software}
The models were trained on a machine with Intel Xeon Silver 4216 CPU @ 2.40GHz having sixty-four cores and 64GBs of RAM. Regarding the implementation of the approaches, the USBCF and the BBCF were implemented in Python. The remaining approaches consider the available implementation in Surprise \citep{Surpriselib-Hug-2020}, 
a Python scikit library for building and analyzing Recommender Systems.

\subsubsection{Datasets}


Experiments were run on two benchmark datasets with different volumes (MovieLens-100k and MovieLens-1M).

The MovieLens-100k\footnote{\url{https://grouplens.org/datasets/movielens/100k/}} dataset \citep{Dataset-Movielens-2016}, one of the most popular publicly available benchmark datasets to serve as domain testing for new CF approaches. This version of the MovieLens' dataset contains 100,000 ordinal ratings (1-5 scale) from 943 users on 1,682 movies. The data is in the form of triplets \textit{\textless user, item, rating\textgreater}. Each user has a minimum of 20 ratings made and a maximum of 737, but the average is 106 ratings per user with a standard deviation of 100 ratings. The sparsity of the MovieLens-100k is 93.6\%, in other words, approximately 6.4\% of the entries are filled.


The MovieLens-1M\footnote{\url{https://grouplens.org/datasets/movielens/1m/}} dataset \citep{Dataset-Movielens-2016} is a bigger version of the MovieLens-100k dataset that contains 1,000,209 ratings of 3,707 movies made by 6,040 users. The sparsity of the MovieLens-1M is 95.54\%.

\subsubsection{Experimental Protocol}
We trained and evaluated all the approaches on the canonical u.base/u.test splits of the MovieLens-100k dataset, used to perform a 5-fold cross-validation with 80\% of the ratings as training data and the remaining 20\% for testing per fold. The same procedure was used for the Movielens-1M dataset.

To measure the rating prediction accuracy of the recommendation algorithms, we used two standard metrics, namely \textit{Mean Absolute Error (MAE)},

\begin{equation}
    MAE = \displaystyle\frac{1}{n}\sum_{j=1}^{n}|\hat{r}_{u_ji_j}-r_{u_ji_j}|,
\end{equation}

\noindent and \textit{Root Mean Squared Error (RMSE)},

\begin{equation}
    RMSE = \displaystyle\sqrt{\frac{1}{n}\sum_{j=1}^{n}\big(\hat{r}_{u_ji_j}-r_{u_ji_j}\big)^2}.
\end{equation}

In this work, beyond the accuracy dimension, we also consider coverage measures for our evaluation, as some approaches provide high quality predictions but for only a small part of the user-item pairs.

\begin{itemize}
    \item \textit{Prediction Coverage.} The percentage of user-item pairs for which a prediction can be made \citep{CF-RecommenderSystemshandbook-Ricci-2015}.
    \item \textit{Item Coverage.} The number of items for which predictions can be formed as a percentage of the total number of items \citep{CF-EvaluateCF-Herlocker-2004}.
\end{itemize}

The optimal hyper-parameters for each model have been estimated using cross-validation on the available training data. Details on the selected hyper-parameters and the experiments reproducibility are discussed in section \ref{subsec:reproducibilityofexperiments}.

\subsection{Performance comparison with popular CF approaches (RQ1)}

In this experiment, we compare the performance of the USBCF against baselines and state-of-the-art rating prediction methods on the two benchmark datasets. The results reported for the USBCF approach use a combination of QUBIC2 biclustering solutions with varying minimum of columns, \textit{minCols}~$\in [3, 5, 7, 10, 15, 20]$, a minimum user-bicluster similarity threshold of 0.25, and an Item-based CF with a neighborhood of 20 items as the CF algorithmic choice. We refer to this model as \textbf{USBCF-IB}.

\subsubsection{Methods}

\begin{itemize}
\item \textit{Baseline Rating Predictor} (\textbf{Bias}) \citep{BaselineCF-bias-Koren2010}: user-item bias rating prediction algorithm, denoted by : $\hat{r}_{i,j}= \mu +  b_i + b_j$, where $\mu$ is the global mean rating, $b_j$ is the item bias, and $b_i$ is the user bias. 
    
    \item \textit{User-based CF} (\textbf{USCF}) \citep{UserBasedCF-Resnick-1994}: user-based CF model with mean-centered cosine similarity as similarity function and weighted-average - taking into account the mean ratings of each user - as aggregation function.
    
    \item \textit{Item-based CF} (\textbf{IBCF}) \citep{ItemBasedCF-Sarwar-2001}: item-based CF model with mean-centered cosine similarity as similarity function and weighted-average - taking into account the mean ratings of each user - as aggregation function.
    
    \item \textit{SVD Model by Simon Funk} (\textbf{FunkSVD}) \citep{MatrixFactCF-funksvd-Simon-2006}: the famous SVD-inspired approach popularized by Simon Funk during the Netflix Prize contest. It uses regularized stochastic gradient descent to train the user-feature and the item-feature matrix.
    
    \item \textit{Singular Value Decomposition with Implicit Feedback} (\textbf{SVD++}) \citep{MatrixFactCF-svd++-koren2008}: 
    an extension of the classic SVD approach that incorporates implicit feedback into the SVD model. 
    
    \item \textit{Non-negative Matrix Factorization} (\textbf{NMF}) \citep{MatrixFactCF-NMF-Xin-2014}:
    non-negative matrix factorization model for rating prediction.
    
    \item \textit{Weighted Bregman Co-Clustering} (\textbf{CoCCF}) \citep{BiclustCF-scalablecf-george-2005}: scalable Collaborative Filtering framework that uses weighted Bregman co-clustering and averages to generate predictions.
    
    \item \textit{Biclustering-based CF} (\textbf{BBCF}) \citep{BCF-impactbiclusteringcf-Singh-2018}: state-of-the-art biclustering-based methodology that uses biclustering as a preprocessing step to scale CF approaches.
    
\end{itemize}

\subsubsection{Results}

\begin{table}[!b]
\centering
\small
\begin{tabular}{@{}l|llll@{}}
\toprule
\multicolumn{1}{c}{Model} &   \multicolumn{1}{c}{MAE} &
  \multicolumn{1}{c}{RMSE} &
  \multicolumn{1}{c}{Coverage (\%)}  &
  \multicolumn{1}{c}{Item Coverage (\%)}
  
  \\ \midrule
\multicolumn{1}{c}{\begin{tabular}[c]{@{}c@{}}Bias\end{tabular}}	& 0.750 $\pm$ 0.007 & 0.946 $\pm$ 0.009 & 100.00 & 100.00\\

\multicolumn{1}{c}{\begin{tabular}[c]{@{}c@{}}UBCF \end{tabular}} & 0.752 $\pm$ 0.005 & 0.959 $\pm$ 0.007 & 99.83  & 97.85\\
\multicolumn{1}{c}{\begin{tabular}[c]{@{}c@{}}IBCF \end{tabular}}& 0.746 $\pm$ 0.004 & 0.952 $\pm$ 0.005 & 99.83  & 97.85\\
\multicolumn{1}{c}{\begin{tabular}[c]{@{}c@{}}FunkSVD  \end{tabular}}	& 0.720 $\pm$ 0.006 & 0.912 $\pm$ 0.006 & 100.00  & 100.00\\
\multicolumn{1}{c}{\begin{tabular}[c]{@{}c@{}}SVD++ \end{tabular}}	 & 0.719 $\pm$ 0.006 & 0.911 $\pm$ 0.007 & 100.00  & 100.00\\
\multicolumn{1}{c}{\begin{tabular}[c]{@{}c@{}}NMF \end{tabular}}	& 0.719 $\pm$ 0.005 & 0.921 $\pm$ 0.006 & 99.83 & 97.85\\
\multicolumn{1}{c}{\begin{tabular}[c]{@{}c@{}}CoCCF\end{tabular}}	 & 0.757 $\pm$ 0.008 & 0.967 $\pm$ 0.010 & 100.00  & 100.00\\
\multicolumn{1}{c}{\begin{tabular}[c]{@{}c@{}}BBCF-IB \end{tabular}}	 & 0.690 $\pm$ 0.007 & 0.886 $\pm$ 0.006 & 42.52  & 42.65\\
\midrule
\multicolumn{1}{c}{\begin{tabular}[c]{@{}c@{}}USBCF-IB \\ (minSim=0.25) \end{tabular}}			 & 0.716 $\pm$ 0.003 & 0.913 $\pm$ 0.004 & 80.03 & 65.14\\ 
\bottomrule
\end{tabular}
\caption{Average predictive results on the MovieLens-100k dataset.}
\label{tab:benchmark_results_mvlens}
\end{table}

Results in \autoref{tab:benchmark_results_mvlens} show that USBCF-IB outperforms the Bias, UBCF, IBCF, and CoCCF models in rating prediction accuracy with statistical significance (at $\alpha$=1E-3), falling behind the BBCF-IB model in terms of accuracy yet significantly surpassing this later model in user and item coverage. Generally, the USBCF-IB approach achieves a rating prediction performance comparable to the matrix factorization-based approaches on the targeted data domain. It is important to note that UBCF, IBCF, BBCF-IB, and USBCF-IB accuracy results could possibly be further improved using a different similarity or aggregation method. For instance, subtracting the item's mean rating instead of the user's mean rating tends to be more effective \citep{CF-rethinkRSresearch-cEkstrand-2011}.

In terms of predictive coverage capability, the compared approaches were able to output a prediction for nearly 100\% of the test pairs, while the USBCF-IB model returns predictions for 80.03\% of the pairs and considers 65.14\% of the overall items, in the absence of the principles introduced along section \ref{coclussec}, nearly doubling the predictive coverage of the BBCF model. Focusing on the USBCF-IB and IBCF models on the Movielens-100k dataset, the results suggest the USBCF-IB approach can improve the IBCF accuracy results with a slight decrease in the coverage capability.

To investigate if the USBCF-IB model produced better predictions for the test pairs it made a prediction, we complemented the USBCF-IB with predictions of alternative approaches for the test pairs the USBCF-IB could not output a prediction. The results in \autoref{tab:usbcf_complemented_coclust_ibmvlens} confirm the USBCF is improving the classic IBCF (k=20) model, and show how the performance of the hybrid approaches vary according to the minimum user-bicluster similarity (minSim) USBCF parameter. Combining USBCF-UB with a CoCCF model also improves the quality of the predictions considerably.

\begin{table}[!t]
\centering
\small
\begin{tabular}{@{}l|ccccc@{}}
\toprule
\multicolumn{1}{c}{Model} &   \multicolumn{1}{c}{MAE} &
  \multicolumn{1}{c}{RMSE} &
  \multicolumn{1}{c}{Coverage (\%)}  &
  \multicolumn{1}{c}{Item Coverage (\%)}
  
  \\ \midrule

 \multicolumn{1}{c}{\begin{tabular}[c]{@{}c@{}}USBCF-IB+IBCF (minSim=0.1)\end{tabular}}			 & 0.742 $\pm$ 0.005 & 0.947 $\pm$ 0.005 & 99.83 & 97.85\\
\multicolumn{1}{c}{\begin{tabular}[c]{@{}c@{}}USBCF-IB+IBCF (minSim=0.2)\end{tabular}}			  & 0.741 $\pm$ 0.004 & 0.946 $\pm$ 0.005 & 99.83 & 97.85\\
\multicolumn{1}{c}{\begin{tabular}[c]{@{}c@{}}USBCF-IB+IBCF (minSim=0.3)\end{tabular}}			& 0.740 $\pm$ 0.004 & 0.945 $\pm$ 0.004 & 99.83 & 97.85\\
\multicolumn{1}{c}{\begin{tabular}[c]{@{}c@{}}USBCF-IB+IBCF (minSim=0.4)\end{tabular}}			  & 0.739 $\pm$ 0.004 & 0.945 $\pm$ 0.005 & 99.83 & 97.85\\
\multicolumn{1}{c}{\begin{tabular}[c]{@{}c@{}}USBCF-IB+IBCF (minSim=0.5)\end{tabular}}			 & 0.740 $\pm$ 0.004 & 0.947 $\pm$ 0.005 & 99.83 & 97.85\\

\midrule

\multicolumn{1}{c}{\begin{tabular}[c]{@{}c@{}}USBCF-IB+CoCCF (minSim=0.1)\end{tabular}}			 & 0.742 $\pm$ 0.004 & 0.947 $\pm$ 0.005 & 100.00 & 100.00\\
\multicolumn{1}{c}{\begin{tabular}[c]{@{}c@{}}USBCF-IB+CoCCF (minSim=0.2)\end{tabular}}			& 0.742 $\pm$ 0.004 & 0.947 $\pm$ 0.005 & 100.00 & 100.00\\
\multicolumn{1}{c}{\begin{tabular}[c]{@{}c@{}}USBCF-IB+CoCCF (minSim=0.3)\end{tabular}}			  & 0.743 $\pm$ 0.005 & 0.950 $\pm$ 0.006 & 100.00 & 100.00\\
\multicolumn{1}{c}{\begin{tabular}[c]{@{}c@{}}USBCF-IB+CoCCF (minSim=0.4)\end{tabular}}			 & 0.745 $\pm$ 0.005 & 0.954 $\pm$ 0.008 & 100.00 & 100.00\\
\multicolumn{1}{c}{\begin{tabular}[c]{@{}c@{}}USBCF-IB+CoCCF (minSim=0.5)\end{tabular}}			  & 0.750 $\pm$ 0.007 & 0.960 $\pm$ 0.010 & 100.00 & 100.00\\

\bottomrule
\end{tabular}
\caption{Predictive accuracy and coverage of USBCF when complemented with other CF approaches on the MovieLens-100k dataset.}
\label{tab:usbcf_complemented_coclust_ibmvlens}
\end{table}

Results on the MovieLens-1M dataset also indicate a competitive predictive accuracy of the proposed approach against the benchmarks methods. \autoref{tab:benchmark_results_mvlens1m} shows that in this data, the USBCF-IB model, with a minimum similarity of 0.25, retrieves an average mean absolute error of 0.686, covering 90.45\% of the user-item pairs. Approaches based on matrix factorization strategies such as NMF were not included as they were not able to converge in due time for adequate parameterization settings considering the size of the dataset. When the complementing USBCF-IB model with the remaining 9.55\% of the predictions, the results in \autoref{tab:usbcf_complemented_coclust_ibmovilens-1m} support the hypothesis that the USBCF-IB model improves the classic IBCF.

\begin{table}[H]
\small
\centering
\begin{tabular}{@{}l|cccc@{}}
\toprule
\multicolumn{1}{c}{Model} &   \multicolumn{1}{c}{MAE} &
  \multicolumn{1}{c}{RMSE} &
  \multicolumn{1}{c}{Coverage (\%)}  &
  \multicolumn{1}{c}{Item Coverage (\%)}
  
  \\ \midrule
\multicolumn{1}{c}{\begin{tabular}[c]{@{}c@{}}Bias\end{tabular}}	 & 0.719 $\pm$ 0.001 & 0.909 $\pm$ 0.001 & 100.00 & 100.00\\

\multicolumn{1}{c}{\begin{tabular}[c]{@{}c@{}}UBCF \end{tabular}}  & 0.735 $\pm$ 0.001 & 0.932 $\pm$ 0.001 & 99.98  & 99.22\\

\multicolumn{1}{c}{\begin{tabular}[c]{@{}c@{}}IBCF\end{tabular}} & 0.704 $\pm$ 0.001 & 0.899 $\pm$ 0.001 & 99.98 & 99.22 \\

\multicolumn{1}{c}{\begin{tabular}[c]{@{}c@{}}CoCCF\end{tabular}}	 & 0.717 $\pm$ 0.001 & 0.915 $\pm$ 0.002 & 100.00  & 100.00\\
\multicolumn{1}{c}{\begin{tabular}[c]{@{}c@{}}BBCF\end{tabular}} 	 & 0.676 $\pm$ 0.001 & 0.866 $\pm$ 0.001 & 71.08 & 61.30\\

\midrule
\multicolumn{1}{c}{\begin{tabular}[c]{@{}c@{}}USBCF-IB (minSim=0.25)\end{tabular}}			 & 0.686 $\pm$ 0.001 & 0.876 $\pm$ 0.001 & 90.45 & 77.21\\
\bottomrule

\end{tabular}
\caption{Average predictive results of baseline and subspace clustering-based approaches on the MovieLens-1M dataset.}
\label{tab:benchmark_results_mvlens1m}
\end{table}

\begin{table}[H]
\resizebox{\textwidth}{!}{%
\centering
\begin{tabular}{@{}l|llll@{}}
\toprule
\multicolumn{1}{c}{Model} &   \multicolumn{1}{c}{MAE} &
  \multicolumn{1}{c}{RMSE} &
  \multicolumn{1}{c}{Coverage (\%)}  &
  \multicolumn{1}{c}{Item Coverage (\%)}
  
  \\ \midrule
  
 \multicolumn{1}{c}{\begin{tabular}[c]{@{}c@{}}USBCF-IB+IBCF (minSim=0.25)\end{tabular}}			 & 0.698 $\pm$ 0.001 & 0.892 $\pm$ 0.001 & 99.98 & 99.22\\

\multicolumn{1}{c}{\begin{tabular}[c]{@{}c@{}}USBCF-IB+CoCCF (minSim=0.25)\end{tabular}} & 0.700 $\pm$ 0.001 & 0.894 $\pm$ 0.001 & 100.00 & 100.00\\

\bottomrule
\end{tabular}
}
\caption{Predictive accuracy and coverage of USBCF complemented with other CF approaches on the MovieLens-1M dataset.}
\label{tab:usbcf_complemented_coclust_ibmovilens-1m}
\end{table}

The testing time of USBCF-IB is computationally efficient, bounded by the lightweighted computational complexity of the subsequent item-based CF step. USBCF-IB training time is also efficient under heuristic biclustering searches and the creation of the personalized dataset (linear on the number of biclusters and their pattern length). Finally, once biclusters are discovered on referenced rating data, USBCF-IB can be applied in an online manner to predict preferences for unseen users.

\subsection{Biclustering and user-bicluster similarity in USBCF (RQ2)}\label{subsec:rq2}

In the second experiment, we study the impact of biclustering search on the USBCF performance using different QUBIC2 parameterizations. We also assess how the user-bicluster similarity affects the size of the personalized datasets produced by the approach, directly impacting the accuracy and coverage results. Accordingly, we varied the parameter of QUBIC2 that controls the shape of biclusters and the USBCF user-bicluster similarity threshold, both 
affecting the number, size and format of the biclusters produced and therefore the properties of the personalized data spaces.

\begin{figure}[H]
        \centering
            \begin{subfigure}{0.49\textwidth}
            \includegraphics[width=\textwidth]{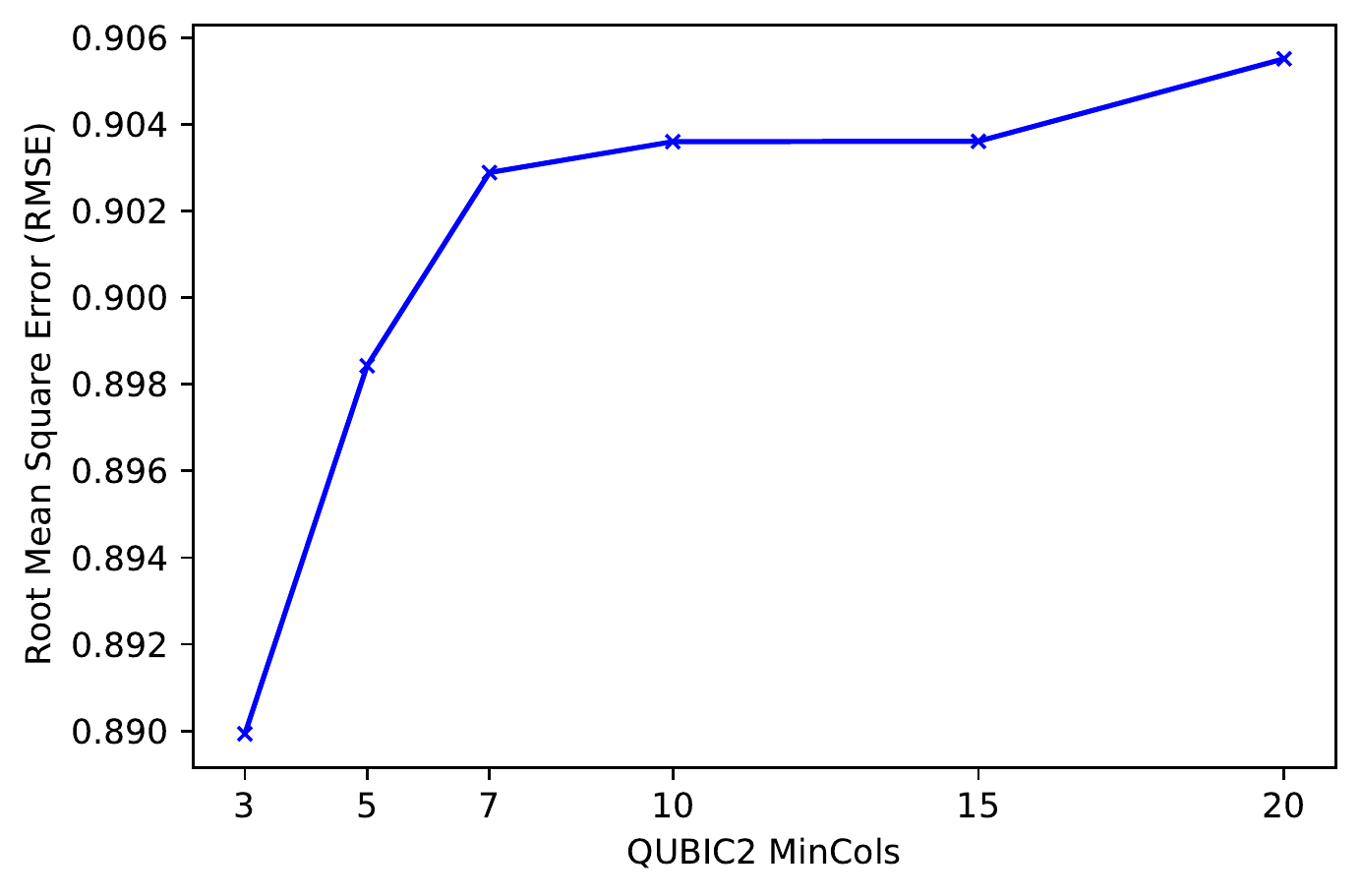}
            \caption{RMSE} 
            \end{subfigure}
           \begin{subfigure}{0.49\textwidth}
            \centering
            \includegraphics[width=\linewidth]{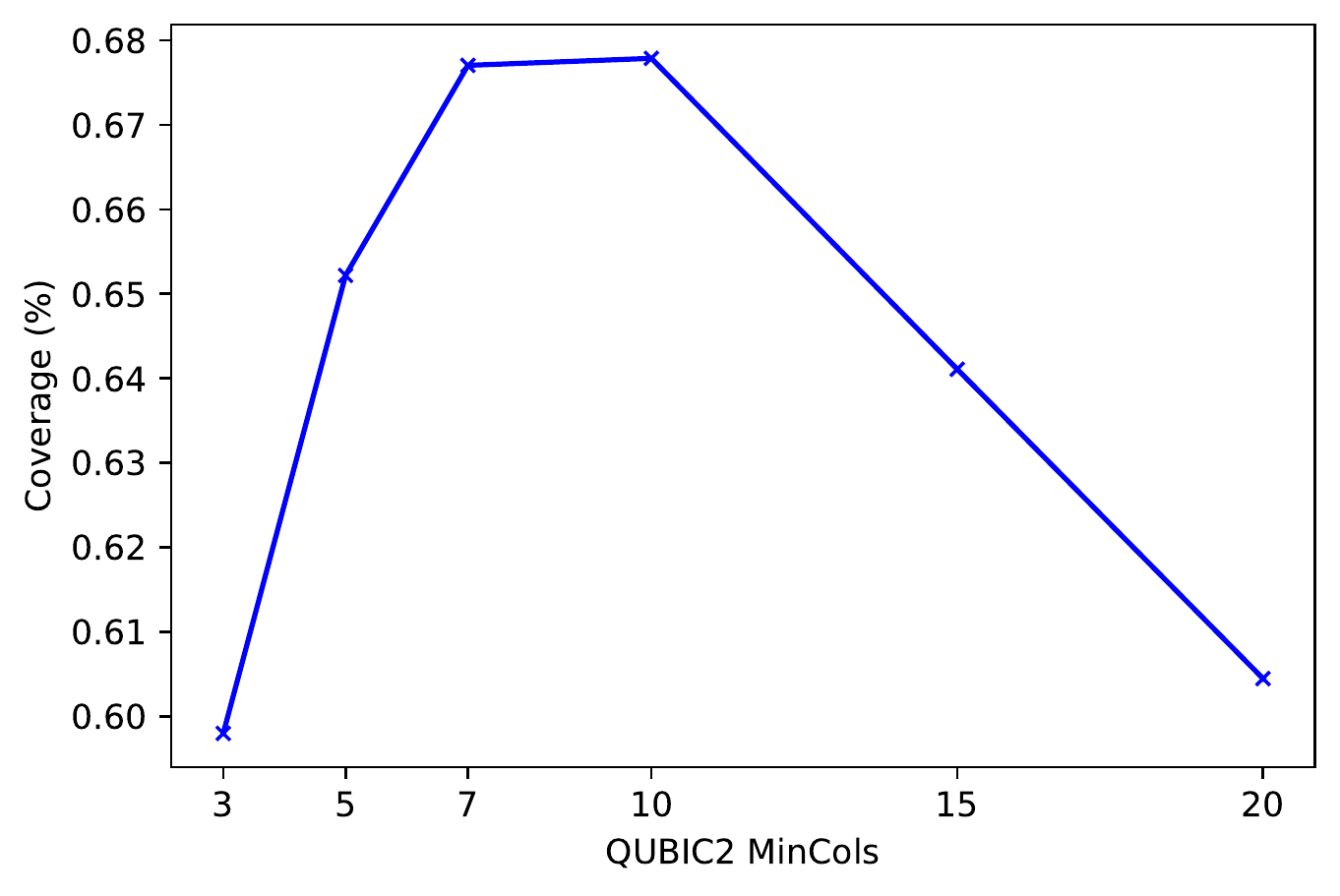}
            \caption{Coverage} 
        \end{subfigure}
    \caption{Effect of the minimum columns per bicluster parameter in the performance.}
    \label{fig:usbcfminsim02_rmse_cov}
    \end{figure}

\autoref{fig:usbcfminsim02_rmse_cov} shows that, for the MovieLens-100k dataset, the average quality of the predictions tends to be worse for biclustering solutions with a larger minimum of columns per bicluster. The coverage capability peaks for biclustering solutions with a minimum of 10 columns per bicluster.

\begin{figure}[H]
        \centering
            \begin{subfigure}{0.49\textwidth}
            \includegraphics[width=\textwidth]{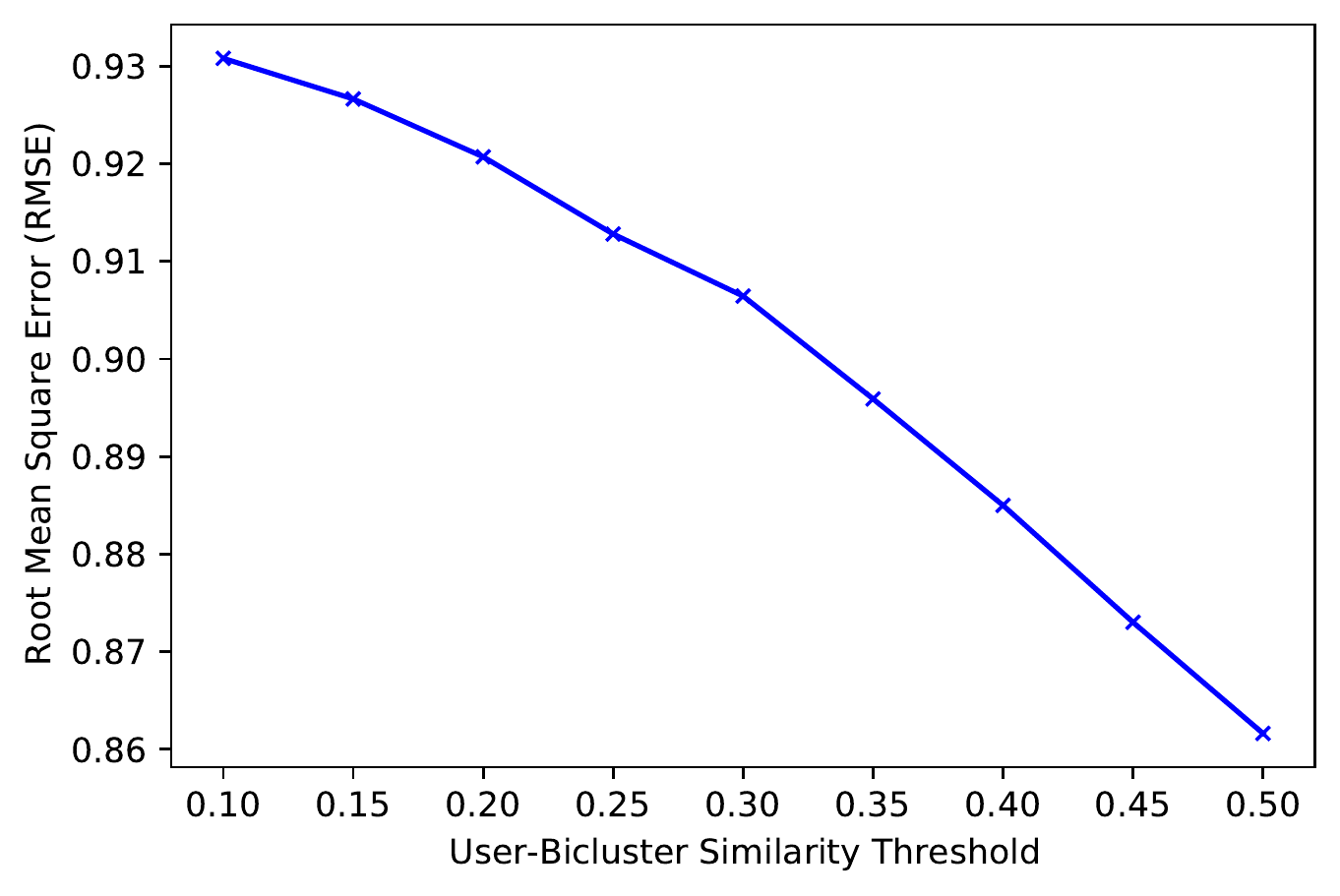}
            \caption{RMSE}
            \label{fig:usbcfcomb_minsim_rmse}
            \end{subfigure}
           \begin{subfigure}{0.49\textwidth}
            \centering
            \includegraphics[width=\linewidth]{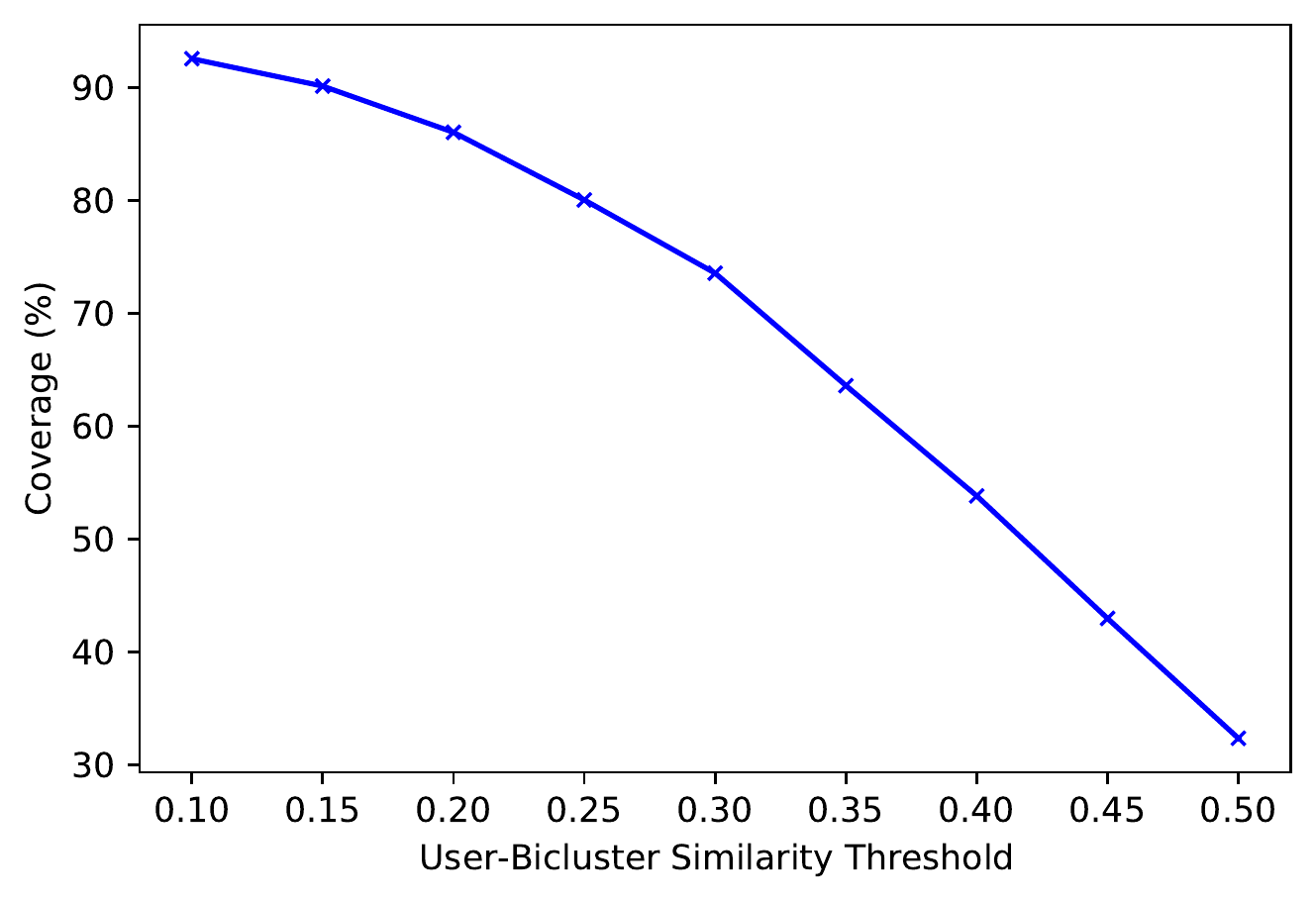}
            \caption{Coverage} 
            \label{fig:usbcfcomb_minsim_cov.pdf}
        \end{subfigure}
    \caption{Sensitivity of USBCF-IB to the user-bicluster similarity threshold parameter. }
    \label{fig:USBCFComb_results_varyingminsim}
    \end{figure}

We also tested the behavior of the approach when using biclusters from multiple biclustering solutions. We ran the biclustering algorithm with the minimum columns parameter in \{3, 5, 7, 10, 15, 20\}, combining the solutions and filtering non-maximal biclusters. The gathered results in \autoref{fig:USBCFComb_results_varyingminsim} against \autoref{fig:usbcfminsim02_rmse_cov} show that the combination of biclustering solutions with distinct properties produces personalized CF models with higher coverage capability and lower RMSE. 

\begin{figure}[H]
        \centering
            \begin{subfigure}{0.49\textwidth}
            \includegraphics[width=\textwidth]{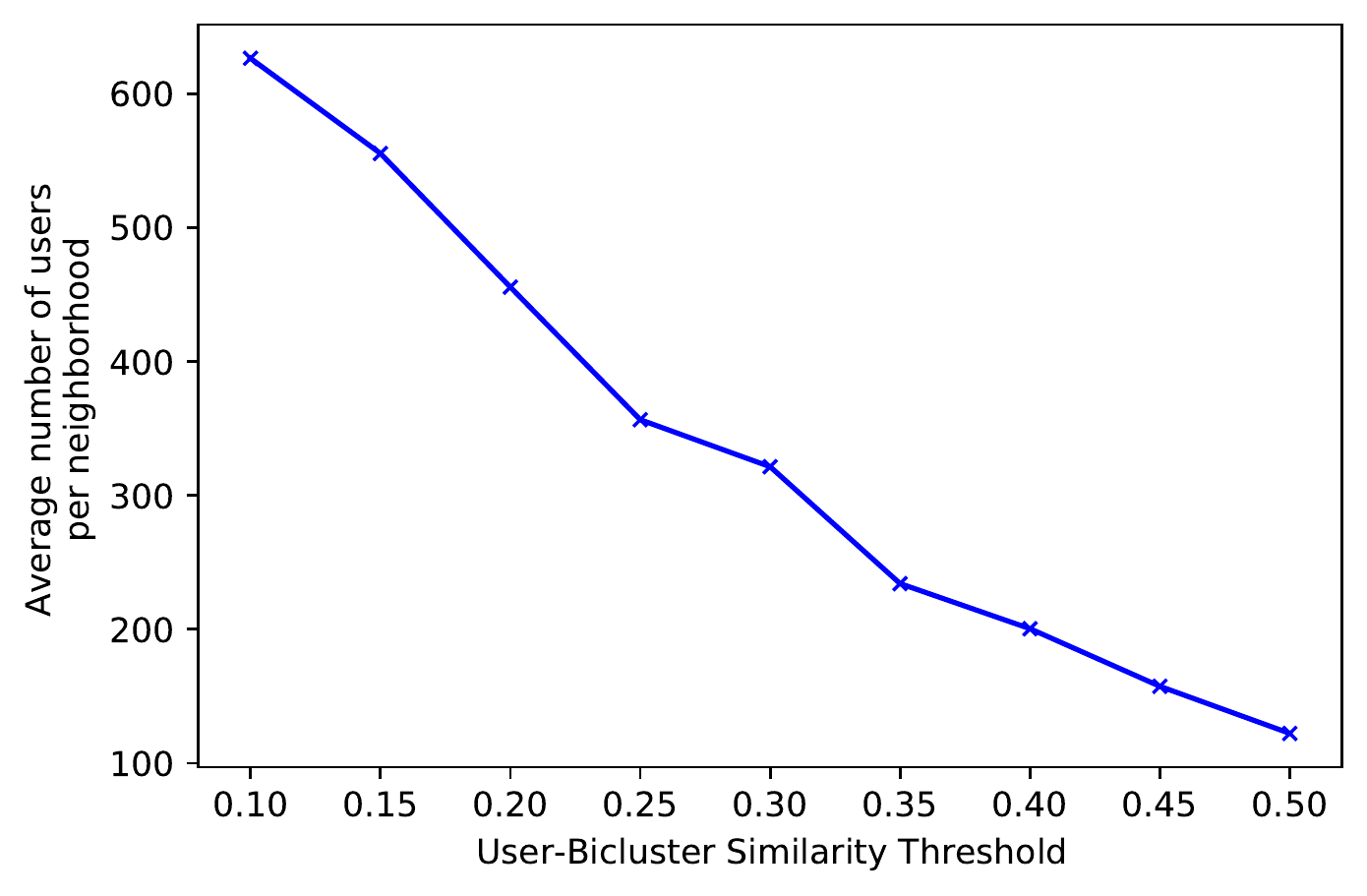}
            \caption{Average number of users per neighborhood.}
            \label{fig:casestudy_bbcf_NNBics_avgrows}
            \end{subfigure}
           \begin{subfigure}{0.49\textwidth}
            \centering
            \includegraphics[width=\linewidth]{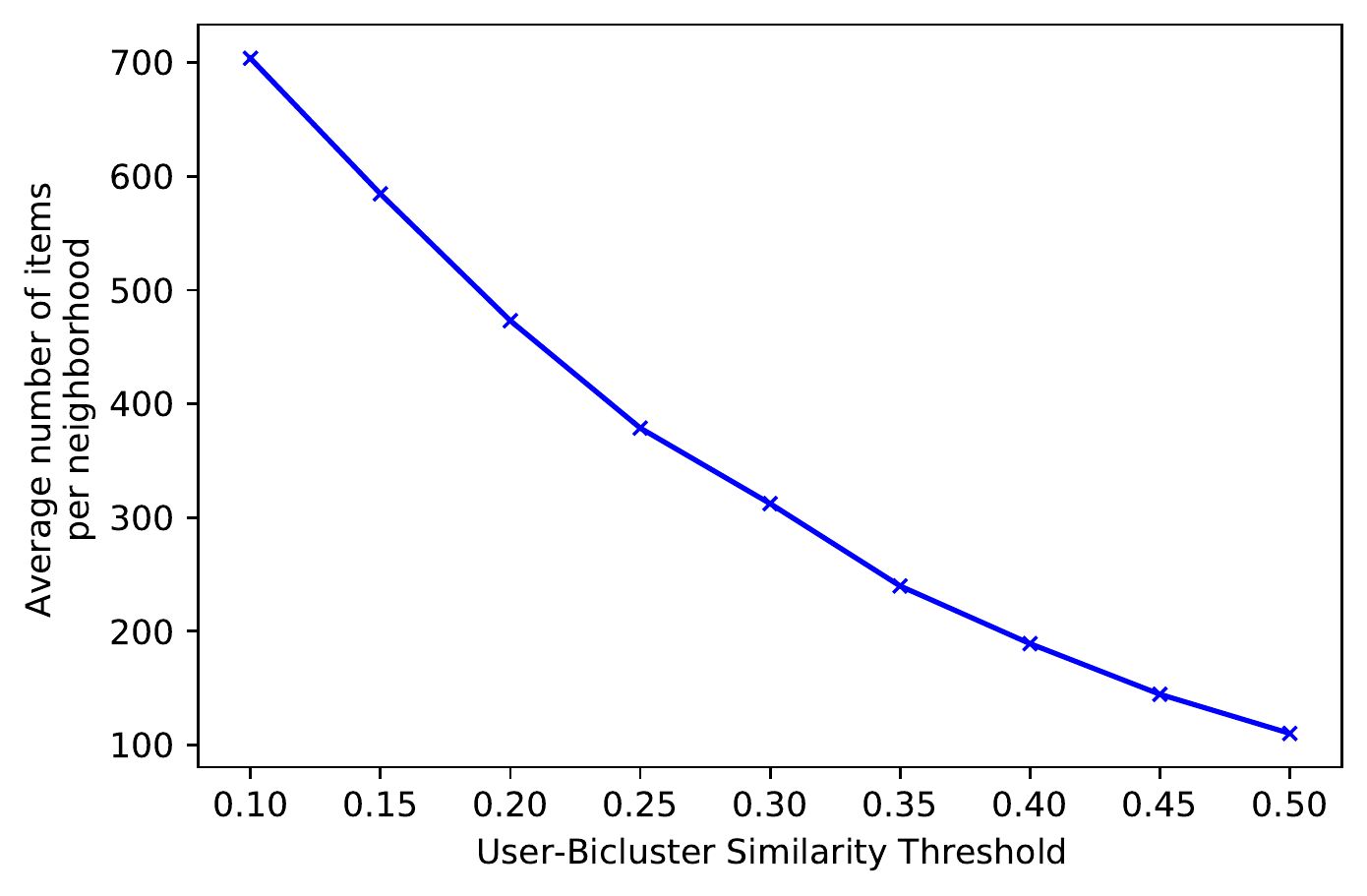}
            \caption{Average number of items per neighborhood.}
            \label{fig:casestudy_bbcf_NNBics_avgcols}
        \end{subfigure}
    \caption{Effect of the minimum user-bicluster similarity threshold on the average size of the personalized data spaces in USBCF. }
    \label{fig:usbcf_min_avgsizes}
    \end{figure}

We finally examined the average size of the personalized datasets when varying the user-bicluster similarity threshold. \autoref{fig:usbcf_min_avgsizes} shows that the USBCF minimum similarity parameter affects the dimensions of the personalized datasets, with lower threshold values creating larger datasets.

\subsection{Performance comparison with Biclustering-based CF (BBCF) (RQ3)}
    
The goal of this third experiment is to compare USBCF against the state-of-the-art biclustering-based methodology, BBCF, introduced by \cite{BCF-impactbiclusteringcf-Singh-2018}. The reported results for BBCF were gathered using an hyperparameterized QUBIC2 biclustering solution, which yielded optimal behavior with a minimum of 15 columns per bicluster. Likewise the BBCF original work, we vary the number of biclusters per neighborhood parameter (\textit{NNBics}).

\begin{figure}[H]
        \centering
            \begin{subfigure}{0.49\textwidth}
            \includegraphics[width=\textwidth]{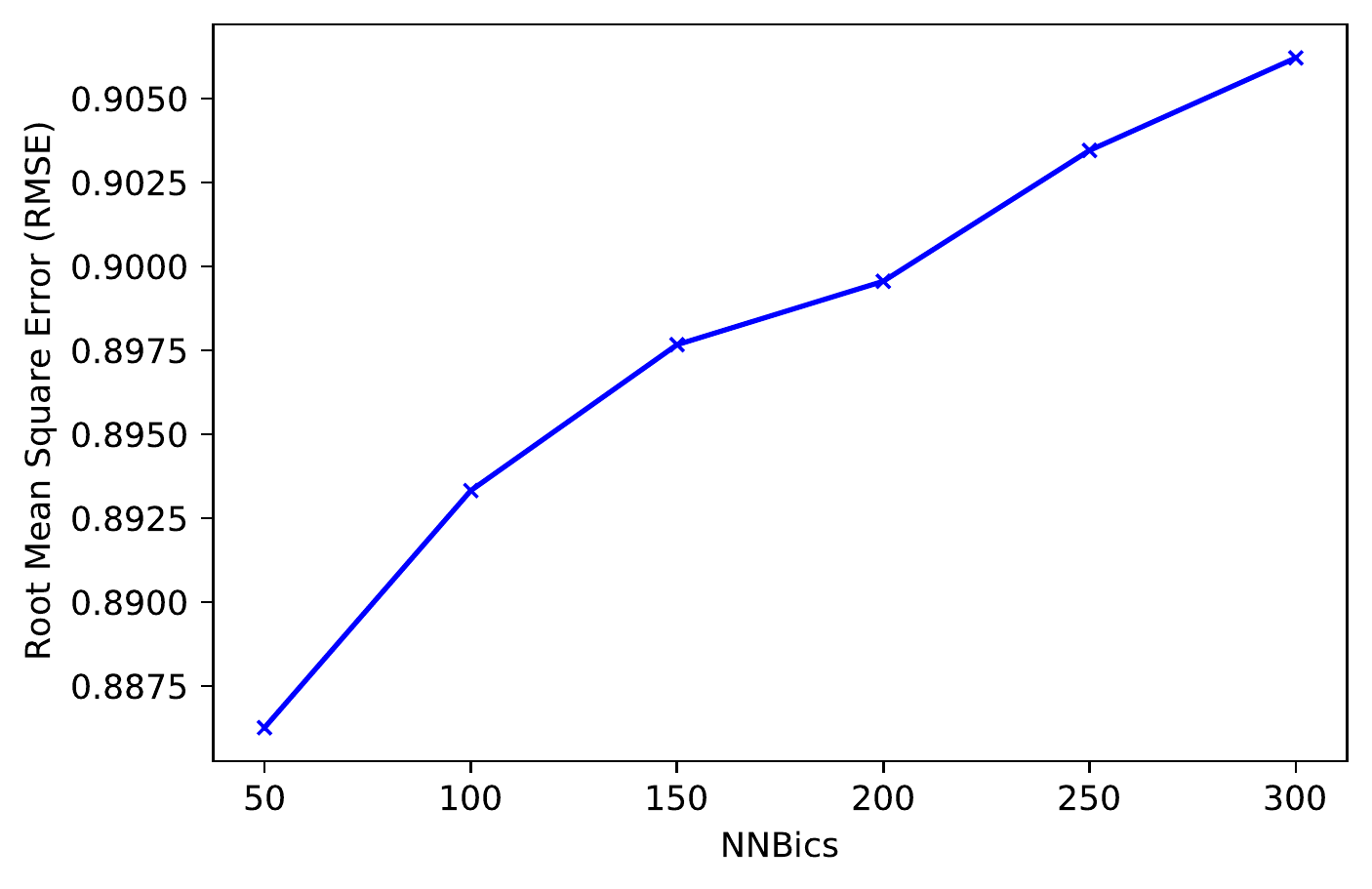}
            \caption{Average RMSE of different biclustering neighborhood sizes.}
            \label{cc}
            \end{subfigure}
           \begin{subfigure}{0.49\textwidth}
            \centering
            \includegraphics[width=\linewidth]{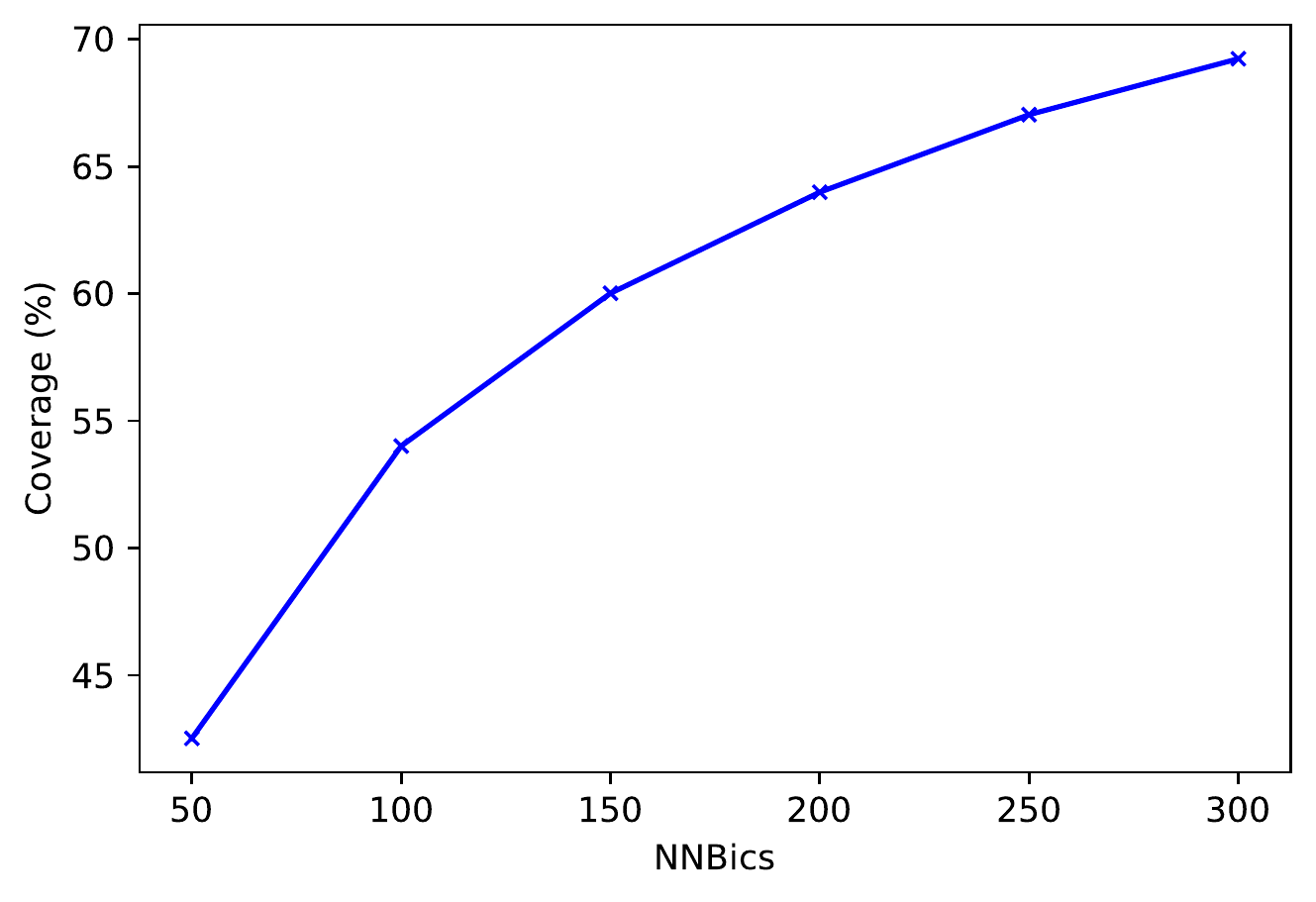}
            \caption{Average coverage of different biclustering neighborhood sizes.}
            \label{aa}
        \end{subfigure}
    \caption{Sensitivity of BBCF to the size of the biclustering neighborhoods. }
    \label{fig:bbcf_nnbrs_rmse_cov}
    \end{figure}

\begin{figure}[H]
        \centering
            \begin{subfigure}{0.49\textwidth}
            \includegraphics[width=\textwidth]{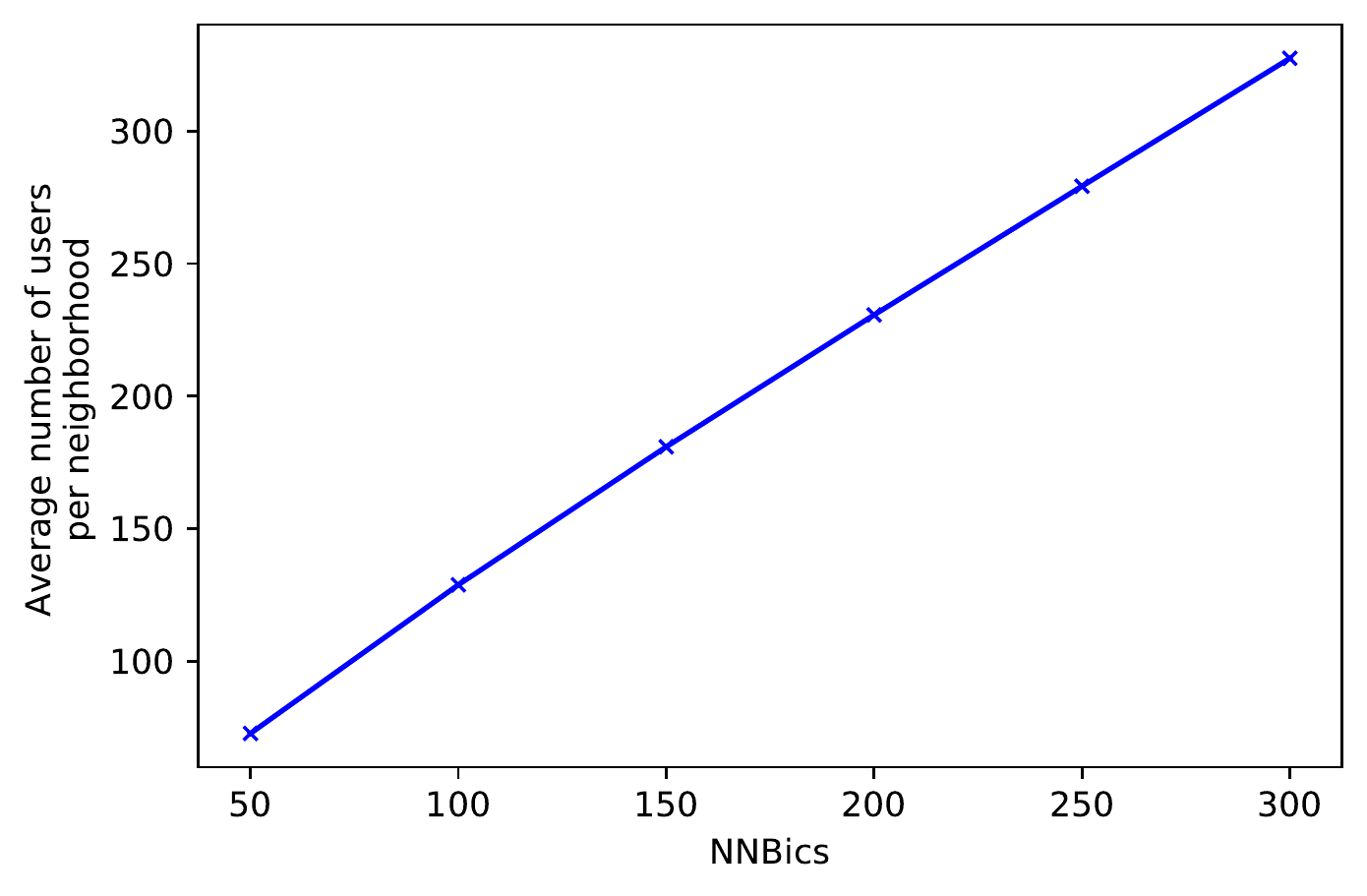}
            \caption{Average number of users per NNBics setup.}
            \end{subfigure}
           \begin{subfigure}{0.49\textwidth}
            \centering
            \includegraphics[width=\linewidth]{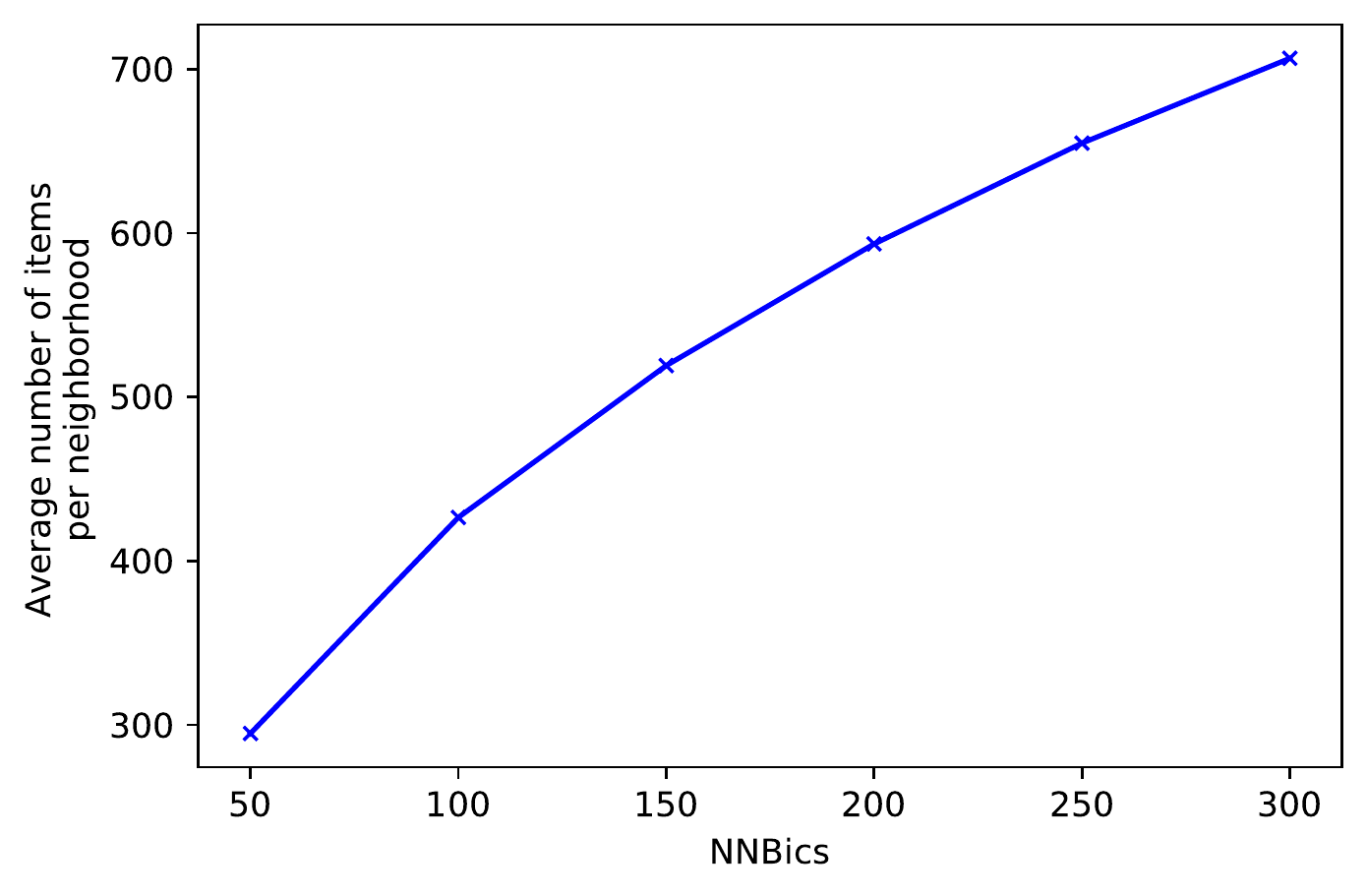}
            \caption{Average number of items per NNBics setup.}
        \end{subfigure}
    \caption{Effect of the number of biclusters in the neighborhood parameter on the average size of the personalized datasets in BBCF.  }
    \label{fig:bbcf_nnbrs_avgsizes}
    \end{figure}

Results in \autoref{fig:bbcf_nnbrs_rmse_cov} show that smaller sets of biclusters per neighborhood are more favorable for the quality of the rating predictions. However, there is a clear RMSE-coverage trade-off, as the model's predictive coverage capability tends to degrade as the number of bicluster per neighborhood decreases. \autoref{fig:bbcf_nnbrs_avgsizes} shows that the size of the personalized datasets increase with larger values for the \textit{NNBics} parameter. These results highlight that, as expected, larger values of \textit{NNBics} lead to more complex personalized models, impacting the scalability of the approach.

Comparing these results with the ones from the previous experiment (\autoref{subsec:rq2}), we conclude that for models with similar coverage, USBCF model surpasses BBCF in prediction error. Moreover, the USBCF approach obtains higher coverage results than BBCF while using smaller personalized datasets.

\subsection{Reproducibility of the Experiments}
\label{subsec:reproducibilityofexperiments}

The source code of USBCF is available at our GitHub repository \footnote{\url{https://git.lasige.di.fc.ul.pt/mmsilva/usbcf-paper}}. The code to reproduce the experiments is also available in the same repository, as well as the obtained results. Regarding the hyper-parameters, we used the following defaults during the experiments: 
\begin{itemize}
    \item UBCF and IBCF: varied maximum number of users/items in the neighborhood (k). We present results for k = 20 as the results show it is an adequate value for the neighborhood size regarding predictive accuracy on both datasets.
    
    \item FunkSVD, SVD++, and NMF: grid search over the three main hyper-parameters for the optimization algorithm: number of features (n\_factors), number of iterations (n\_epochs) and regularization term (reg). 
    For the MovieLens-100k data, the best hyper-parameters were n\_factors = 100, n\_epochs = 100, reg = 0.1 for SVD;  n\_factors = 30, n\_epochs = 50, reg = 0.1 for SVD++; and  n\_factors = 200, n\_epochs = 100, reg = 0.05 for NMF. The reported results use these optimal hyper-parameters estimates. As for the MovieLens-1M dataset, the best hyper-parameters were not attained as the approaches did not converge in due time.
    
    \item BBCF-IB: Varied the hyper-parameter number of nearest biclusters (nnbrs) from 50 to 300 for the Movielens-100k dataset and from 50 to 1000 for the Movielens-1M dataset, both using an Item-Based approach (k = 20) as CF algorithm. We also tested the minimum number of columns per bicluster (minCols) QUBIC2 parameter from 3 to 20. The best hyper-parameters considering predictive and coverage results were nnbrs = 50 and minCols = 15 for the Movielens-100k dataset; and nnbrs = 500 and minCols = 15 for the Movielens-1M dataset.
    
    \item USBCF-IB: Varied the minimum user-bicluster similarity threshold (minSim) parameter from 0.1 to 0.5, and used an Item-Based approach (k = 20) as CF algorithm. The best minSim considering predictive and coverage results were minSim = 0.25 for both datasets.
    
\end{itemize}

\section{Conclusions and Future Work}
\label{conclusions}

In this paper, we propose a user-specific CF approach that uses biclustering as a sub-step to improve traditional CF methods. 
Principles for superior CF were introduced, including superior matching of active user preferences against preference patterns to account for preference diversity; formation of user-personalized data spaces to account for preference locality and sparsity; and secondary rating estimates from coclustering structures to yield high predictive coverage.

Experimental results on the popular MovieLens-100k and Movielens-1M benchmark datasets 
against state-of-the-art CF methods 
show that USBCF yields competitive rating prediction errors, significantly improving classic memory-based CF approaches by alleviating the sparsity of the datasets. Moreover, the proposed approach outperforms the state-of-the-art biclustering-based approach (BBCF) in prediction quality while successfully overcoming BBCF's major limitation, the predictive coverage capability. 

This work displays the potentialities of the biclustering technique for recommendation purposes. Accordingly, we highlight potentially relevant research avenues for future work:

\begin{itemize}

    
    \item explore the impact that different coherence assumptions, degrees of noise tolerance, and biclustering searches yield on the USBCF approach. 
    As USBCF can be parameterized with any biclustering algorithm with minor modifications to the overall methodology, we aim at assessing the role of order-preserving preference patterns in CF recurring to BicPAMS \citep{BC-BicPams-Rui-2017}, quality relaxations (percentage of noisy and missing elements) \citep{henriques2014bicpam,BC-QUBIC2-2019}, and iterative search-and-masking procedures \citep{BC-BicPams-Rui-2017} to guarantee a better coverage of the original rating data space; 

    \item incorporate available background knowledge associated with the profile of users, as well as metadata related to the composition and characteristics of items, to guide the USBCF approach using knowledge-enriched biclustering searches \citep{BC-BIC2PAM-Rui-2016}; 

    \item extend USBCF to online settings. USBCF is as-is readily applicable to new users. Still, to handle preference drift along time, the underlying biclustering solutions used to create the personalized data spaces must be updated with newly arriving user-item ratings. To prevent the need to continuously retrain USBCF, we aim at developing an incremental version of USBCF based on updatable biclustering searches where new items and users can be added/remove to the subspaces in order to explore the inherently temporal context of user-item ratings. 
\end{itemize}

\begin{acks}
\noindent This work was supported by the Fundação para a Ciência e a Tecnologia (FCT) under LASIGE strategic project (UIDB/00408/2020 and UIDP/00408/2020), iCare4U (PTDC/EME-SIS/31474/2017), WISDOM (DSAIPA/DS/0089/2018), ILU (DSAIPA/ DS/0111/2018) and INESC-ID (UIDB/50021/2020).
\end{acks}

\bibliographystyle{ACM-Reference-Format}
\bibliography{biclustering, biclusteringCF, collaborativefiltering}


\begin{thebibliography}{55}


\ifx \showCODEN    \undefined \def \showCODEN     #1{\unskip}     \fi
\ifx \showDOI      \undefined \def \showDOI       #1{#1}\fi
\ifx \showISBNx    \undefined \def \showISBNx     #1{\unskip}     \fi
\ifx \showISBNxiii \undefined \def \showISBNxiii  #1{\unskip}     \fi
\ifx \showISSN     \undefined \def \showISSN      #1{\unskip}     \fi
\ifx \showLCCN     \undefined \def \showLCCN      #1{\unskip}     \fi
\ifx \shownote     \undefined \def \shownote      #1{#1}          \fi
\ifx \showarticletitle \undefined \def \showarticletitle #1{#1}   \fi
\ifx \showURL      \undefined \def \showURL       {\relax}        \fi
\providecommand\bibfield[2]{#2}
\providecommand\bibinfo[2]{#2}
\providecommand\natexlab[1]{#1}
\providecommand\showeprint[2][]{arXiv:#2}

\bibitem[\protect\citeauthoryear{Alqadah and Bhatnagar}{Alqadah and
  Bhatnagar}{2011}]%
        {BC-biclustneighborhoodtopn-bicsimilarity}
\bibfield{author}{\bibinfo{person}{Faris Alqadah} {and} \bibinfo{person}{Raj
  Bhatnagar}.} \bibinfo{year}{2011}\natexlab{}.
\newblock \showarticletitle{Similarity measures in formal concept analysis}.
\newblock \bibinfo{journal}{\emph{Ann. Math. Artif. Intell.}}
  \bibinfo{volume}{61}, \bibinfo{number}{3} (\bibinfo{year}{2011}),
  \bibinfo{pages}{245--256}.
\newblock
\urldef\tempurl%
\url{https://doi.org/10.1007/s10472-011-9257-7}
\showDOI{\tempurl}


\bibitem[\protect\citeauthoryear{Alqadah, Reddy, Hu, and Alqadah}{Alqadah
  et~al\mbox{.}}{2015}]%
        {BCF-biclustneighborhoodtopn-alqadah-2015}
\bibfield{author}{\bibinfo{person}{Faris Alqadah}, \bibinfo{person}{Chandan~K.
  Reddy}, \bibinfo{person}{Junling Hu}, {and} \bibinfo{person}{Hatim~F.
  Alqadah}.} \bibinfo{year}{2015}\natexlab{}.
\newblock \showarticletitle{Biclustering neighborhood-based collaborative
  filtering method for top-n recommender systems}.
\newblock \bibinfo{journal}{\emph{Knowl. Inf. Syst.}} \bibinfo{volume}{44},
  \bibinfo{number}{2} (\bibinfo{year}{2015}), \bibinfo{pages}{475--491}.
\newblock
\urldef\tempurl%
\url{https://doi.org/10.1007/s10115-014-0771-x}
\showDOI{\tempurl}


\bibitem[\protect\citeauthoryear{Banerjee, Dhillon, Ghosh, Merugu, and
  Modha}{Banerjee et~al\mbox{.}}{2007}]%
        {BC-Bregmancoclust-2004}
\bibfield{author}{\bibinfo{person}{Arindam Banerjee},
  \bibinfo{person}{Inderjit~S. Dhillon}, \bibinfo{person}{Joydeep Ghosh},
  \bibinfo{person}{Srujana Merugu}, {and} \bibinfo{person}{Dharmendra~S.
  Modha}.} \bibinfo{year}{2007}\natexlab{}.
\newblock \showarticletitle{A Generalized Maximum Entropy Approach to Bregman
  Co-clustering and Matrix Approximation}.
\newblock \bibinfo{journal}{\emph{J. Mach. Learn. Res.}}  \bibinfo{volume}{8}
  (\bibinfo{year}{2007}), \bibinfo{pages}{1919--1986}.
\newblock
\urldef\tempurl%
\url{http://dl.acm.org/citation.cfm?id=1314563}
\showURL{%
\tempurl}


\bibitem[\protect\citeauthoryear{Bennett and Lanning}{Bennett and
  Lanning}{2007}]%
        {CF-Dataset-Netflix}
\bibfield{author}{\bibinfo{person}{J. Bennett} {and} \bibinfo{person}{S.
  Lanning}.} \bibinfo{year}{2007}\natexlab{}.
\newblock \showarticletitle{The Netflix Prize}. In
  \bibinfo{booktitle}{\emph{Proceedings of the KDD Cup Workshop 2007}}.
  \bibinfo{publisher}{ACM}, \bibinfo{address}{New York}, \bibinfo{pages}{3--6}.
\newblock
\urldef\tempurl%
\url{http://www.cs.uic.edu/~liub/KDD-cup-2007/NetflixPrize-description.pdf}
\showURL{%
\tempurl}


\bibitem[\protect\citeauthoryear{C~.~Madeira, Cacho~Teixeira, Sá-Correia, and
  Oliveira}{C~.~Madeira et~al\mbox{.}}{2010}]%
        {BC-geneexpression-Sara-2010}
\bibfield{author}{\bibinfo{person}{Sara C~.~Madeira}, \bibinfo{person}{Miguel
  Cacho~Teixeira}, \bibinfo{person}{Isabel Sá-Correia}, {and}
  \bibinfo{person}{Arlindo Oliveira}.} \bibinfo{year}{2010}\natexlab{}.
\newblock \showarticletitle{Identification of Regulatory Modules in Time Series
  Gene Expression Data Using a Linear Time Biclustering Algorithm}.
\newblock \bibinfo{journal}{\emph{IEEE/ACM transactions on computational
  biology and bioinformatics / IEEE, ACM}}  \bibinfo{volume}{7}
  (\bibinfo{date}{04} \bibinfo{year}{2010}), \bibinfo{pages}{153--65}.
\newblock
\urldef\tempurl%
\url{https://doi.org/10.1109/TCBB.2008.34}
\showDOI{\tempurl}


\bibitem[\protect\citeauthoryear{Cheng and Church}{Cheng and Church}{2000a}]%
        {BC-geneexpression-Church-2000}
\bibfield{author}{\bibinfo{person}{Yizong Cheng} {and}
  \bibinfo{person}{George~M. Church}.} \bibinfo{year}{2000}\natexlab{a}.
\newblock \showarticletitle{Biclustering of Expression Data}. In
  \bibinfo{booktitle}{\emph{Proceedings of the Eighth International Conference
  on Intelligent Systems for Molecular Biology, August 19-23, 2000, La Jolla /
  San Diego, CA, {USA}}}, \bibfield{editor}{\bibinfo{person}{Philip~E. Bourne},
  \bibinfo{person}{Michael Gribskov}, \bibinfo{person}{Russ~B. Altman},
  \bibinfo{person}{Nancy Jensen}, \bibinfo{person}{Debra~A. Hope},
  \bibinfo{person}{Thomas Lengauer}, \bibinfo{person}{Julie~C. Mitchell},
  \bibinfo{person}{Eric~D. Scheeff}, \bibinfo{person}{Chris Smith},
  \bibinfo{person}{Shawn Strande}, {and} \bibinfo{person}{Helge Weissig}}
  (Eds.). \bibinfo{publisher}{{AAAI}}, \bibinfo{pages}{93--103}.
\newblock
\urldef\tempurl%
\url{http://www.aaai.org/Library/ISMB/2000/ismb00-010.php}
\showURL{%
\tempurl}


\bibitem[\protect\citeauthoryear{Cheng and Church}{Cheng and Church}{2000b}]%
        {BC-ChengChurch-2000}
\bibfield{author}{\bibinfo{person}{Yizong Cheng} {and}
  \bibinfo{person}{George~M Church}.} \bibinfo{year}{2000}\natexlab{b}.
\newblock \showarticletitle{Biclustering of expression data.}. In
  \bibinfo{booktitle}{\emph{Ismb}}, Vol.~\bibinfo{volume}{8}.
  \bibinfo{pages}{93--103}.
\newblock


\bibitem[\protect\citeauthoryear{Coelho, de~Fran{\c{c}}a, and Zuben}{Coelho
  et~al\mbox{.}}{2008}]%
        {BC-MOMainet-Coelho2008}
\bibfield{author}{\bibinfo{person}{Guilherme~Palermo Coelho},
  \bibinfo{person}{Fabr{\'{\i}}cio~Olivetti de Fran{\c{c}}a}, {and}
  \bibinfo{person}{Fernando J.~Von Zuben}.} \bibinfo{year}{2008}\natexlab{}.
\newblock \showarticletitle{A Multi-Objective Multipopulation Approach for
  Biclustering}. In \bibinfo{booktitle}{\emph{Artificial Immune Systems, 7th
  International Conference, {ICARIS} 2008, Phuket, Thailand, August 10-13,
  2008. Proceedings}} \emph{(\bibinfo{series}{Lecture Notes in Computer
  Science}, Vol.~\bibinfo{volume}{5132})},
  \bibfield{editor}{\bibinfo{person}{Peter~J. Bentley}, \bibinfo{person}{Doheon
  Lee}, {and} \bibinfo{person}{Sungwon Jung}} (Eds.).
  \bibinfo{publisher}{Springer}, \bibinfo{pages}{71--82}.
\newblock
\urldef\tempurl%
\url{https://doi.org/10.1007/978-3-540-85072-4\_7}
\showDOI{\tempurl}


\bibitem[\protect\citeauthoryear{de~Castro, de~Fran{\c{c}}a, Ferreira, and
  Zuben}{de~Castro et~al\mbox{.}}{2007}]%
        {BCF-applybiclustcf-Castro-2007}
\bibfield{author}{\bibinfo{person}{Pablo A.~D. de Castro},
  \bibinfo{person}{Fabr{\'{\i}}cio~Olivetti de Fran{\c{c}}a},
  \bibinfo{person}{Hamilton~M. Ferreira}, {and} \bibinfo{person}{Fernando
  J.~Von Zuben}.} \bibinfo{year}{2007}\natexlab{}.
\newblock \showarticletitle{Applying Biclustering to Perform Collaborative
  Filtering}. In \bibinfo{booktitle}{\emph{Seventh International Conference on
  Intelligent Systems Design and Applications, {ISDA} 2007, Rio de Janeiro,
  Brazil, October 20-24, 2007}}, \bibfield{editor}{\bibinfo{person}{Luiza
  de~Macedo~Mourelle}, \bibinfo{person}{Nadia Nedjah}, \bibinfo{person}{Janusz
  Kacprzyk}, {and} \bibinfo{person}{Ajith Abraham}} (Eds.).
  \bibinfo{publisher}{{IEEE} Computer Society}, \bibinfo{pages}{421--426}.
\newblock
\urldef\tempurl%
\url{https://doi.org/10.1109/ISDA.2007.91}
\showDOI{\tempurl}


\bibitem[\protect\citeauthoryear{De~Fran{\c{c}}a, Coelho, and
  Von~Zuben}{De~Fran{\c{c}}a et~al\mbox{.}}{2009}]%
        {BCF-franca-2009}
\bibfield{author}{\bibinfo{person}{Fabricio~Olivetti De~Fran{\c{c}}a},
  \bibinfo{person}{Guilherme~Palermo Coelho}, {and} \bibinfo{person}{Fernando~J
  Von~Zuben}.} \bibinfo{year}{2009}\natexlab{}.
\newblock \showarticletitle{Coherent recommendations using biclustering}. In
  \bibinfo{booktitle}{\emph{Proc. of the XXX Congresso Ibero-Latino-Americano
  de M{\'e}todos Computacionais em Engenharia (CILAMCE)}}.
  \bibinfo{pages}{1--15}.
\newblock


\bibitem[\protect\citeauthoryear{Dhillon, Mallela, and Modha}{Dhillon
  et~al\mbox{.}}{2003}]%
        {BC-ITCC-Dhillon-2003}
\bibfield{author}{\bibinfo{person}{Inderjit~S. Dhillon},
  \bibinfo{person}{Subramanyam Mallela}, {and} \bibinfo{person}{Dharmendra~S.
  Modha}.} \bibinfo{year}{2003}\natexlab{}.
\newblock \showarticletitle{Information-theoretic co-clustering}. In
  \bibinfo{booktitle}{\emph{Proceedings of the Ninth {ACM} {SIGKDD}
  International Conference on Knowledge Discovery and Data Mining, Washington,
  DC, USA, August 24 - 27, 2003}}, \bibfield{editor}{\bibinfo{person}{Lise
  Getoor}, \bibinfo{person}{Ted~E. Senator}, \bibinfo{person}{Pedro~M.
  Domingos}, {and} \bibinfo{person}{Christos Faloutsos}} (Eds.).
  \bibinfo{publisher}{{ACM}}, \bibinfo{pages}{89--98}.
\newblock
\urldef\tempurl%
\url{https://doi.org/10.1145/956750.956764}
\showDOI{\tempurl}


\bibitem[\protect\citeauthoryear{Ekstrand, Ludwig, Konstan, and Riedl}{Ekstrand
  et~al\mbox{.}}{2011a}]%
        {CF-rethinkRSresearch-cEkstrand-2011}
\bibfield{author}{\bibinfo{person}{Michael~D. Ekstrand},
  \bibinfo{person}{Michael Ludwig}, \bibinfo{person}{Joseph~A. Konstan}, {and}
  \bibinfo{person}{John Riedl}.} \bibinfo{year}{2011}\natexlab{a}.
\newblock \showarticletitle{Rethinking the recommender research ecosystem:
  reproducibility, openness, and LensKit}. In
  \bibinfo{booktitle}{\emph{Proceedings of the 2011 {ACM} Conference on
  Recommender Systems, RecSys 2011, Chicago, IL, USA, October 23-27, 2011}},
  \bibfield{editor}{\bibinfo{person}{Bamshad Mobasher},
  \bibinfo{person}{Robin~D. Burke}, \bibinfo{person}{Dietmar Jannach}, {and}
  \bibinfo{person}{Gediminas Adomavicius}} (Eds.). \bibinfo{publisher}{{ACM}},
  \bibinfo{pages}{133--140}.
\newblock
\urldef\tempurl%
\url{https://doi.org/10.1145/2043932.2043958}
\showDOI{\tempurl}


\bibitem[\protect\citeauthoryear{Ekstrand, Riedl, and Konstan}{Ekstrand
  et~al\mbox{.}}{2011b}]%
        {CF-Survey-Ekstrand-2011}
\bibfield{author}{\bibinfo{person}{Michael~D. Ekstrand}, \bibinfo{person}{John
  Riedl}, {and} \bibinfo{person}{Joseph~A. Konstan}.}
  \bibinfo{year}{2011}\natexlab{b}.
\newblock \showarticletitle{Collaborative Filtering Recommender Systems}.
\newblock \bibinfo{journal}{\emph{Found. Trends Hum. Comput. Interact.}}
  \bibinfo{volume}{4}, \bibinfo{number}{2} (\bibinfo{year}{2011}),
  \bibinfo{pages}{175--243}.
\newblock
\urldef\tempurl%
\url{https://doi.org/10.1561/1100000009}
\showDOI{\tempurl}


\bibitem[\protect\citeauthoryear{El{-}Nabarawy, II, and Abdelbar}{El{-}Nabarawy
  et~al\mbox{.}}{2016}]%
        {BCF-artmap-elnabarawy-2016}
\bibfield{author}{\bibinfo{person}{Islam El{-}Nabarawy},
  \bibinfo{person}{Donald C.~Wunsch II}, {and} \bibinfo{person}{Ashraf~M.
  Abdelbar}.} \bibinfo{year}{2016}\natexlab{}.
\newblock \showarticletitle{Biclustering {ARTMAP} collaborative filtering
  recommender system}. In \bibinfo{booktitle}{\emph{2016 International Joint
  Conference on Neural Networks, {IJCNN} 2016, Vancouver, BC, Canada, July
  24-29, 2016}}. \bibinfo{publisher}{{IEEE}}, \bibinfo{pages}{2986--2991}.
\newblock
\urldef\tempurl%
\url{https://doi.org/10.1109/IJCNN.2016.7727578}
\showDOI{\tempurl}


\bibitem[\protect\citeauthoryear{Feng, Zhao, and Zhou}{Feng
  et~al\mbox{.}}{2020}]%
        {BCF-improvetopnrec-Feng-2020}
\bibfield{author}{\bibinfo{person}{Liang Feng}, \bibinfo{person}{Qianchuan
  Zhao}, {and} \bibinfo{person}{Cangqi Zhou}.} \bibinfo{year}{2020}\natexlab{}.
\newblock \showarticletitle{Improving performances of Top-\emph{N}
  recommendations with co-clustering method}.
\newblock \bibinfo{journal}{\emph{Expert Syst. Appl.}}  \bibinfo{volume}{143}
  (\bibinfo{year}{2020}).
\newblock
\urldef\tempurl%
\url{https://doi.org/10.1016/j.eswa.2019.113078}
\showDOI{\tempurl}


\bibitem[\protect\citeauthoryear{Funk}{Funk}{2006}]%
        {MatrixFactCF-funksvd-Simon-2006}
\bibfield{author}{\bibinfo{person}{Simon Funk}.}
  \bibinfo{year}{2006}\natexlab{}.
\newblock \bibinfo{title}{Netflix Update: Try This At Home}.
\newblock \bibinfo{howpublished}{Retrieved from
  https://sifter.org/~simon/journal/20061211.html}.
\newblock
\newblock
\shownote{Accessed: February 11, 2021.}


\bibitem[\protect\citeauthoryear{Ganter and Wille}{Ganter and Wille}{1999}]%
        {BC-biclustneighborhoodtopn-FCA-BOOK}
\bibfield{author}{\bibinfo{person}{Bernhard Ganter} {and}
  \bibinfo{person}{Rudolf Wille}.} \bibinfo{year}{1999}\natexlab{}.
\newblock \bibinfo{booktitle}{\emph{Formal Concept Analysis - Mathematical
  Foundations}}.
\newblock \bibinfo{publisher}{Springer}.
\newblock
\showISBNx{978-3-540-62771-5}
\urldef\tempurl%
\url{https://doi.org/10.1007/978-3-642-59830-2}
\showDOI{\tempurl}


\bibitem[\protect\citeauthoryear{George and Merugu}{George and Merugu}{2005}]%
        {BiclustCF-scalablecf-george-2005}
\bibfield{author}{\bibinfo{person}{Thomas George} {and}
  \bibinfo{person}{Srujana Merugu}.} \bibinfo{year}{2005}\natexlab{}.
\newblock \showarticletitle{A Scalable Collaborative Filtering Framework Based
  on Co-Clustering}. In \bibinfo{booktitle}{\emph{Proceedings of the 5th {IEEE}
  International Conference on Data Mining {(ICDM} 2005), 27-30 November 2005,
  Houston, Texas, {USA}}}. \bibinfo{publisher}{{IEEE} Computer Society},
  \bibinfo{pages}{625--628}.
\newblock
\urldef\tempurl%
\url{https://doi.org/10.1109/ICDM.2005.14}
\showDOI{\tempurl}


\bibitem[\protect\citeauthoryear{Goldberg, Nichols, Oki, and Terry}{Goldberg
  et~al\mbox{.}}{1992}]%
        {CF-UserBasedCF-Goldberg-1992}
\bibfield{author}{\bibinfo{person}{David Goldberg}, \bibinfo{person}{David~A.
  Nichols}, \bibinfo{person}{Brian~M. Oki}, {and} \bibinfo{person}{Douglas~B.
  Terry}.} \bibinfo{year}{1992}\natexlab{}.
\newblock \showarticletitle{Using Collaborative Filtering to Weave an
  Information Tapestry}.
\newblock \bibinfo{journal}{\emph{Commun. {ACM}}} \bibinfo{volume}{35},
  \bibinfo{number}{12} (\bibinfo{year}{1992}), \bibinfo{pages}{61--70}.
\newblock
\urldef\tempurl%
\url{https://doi.org/10.1145/138859.138867}
\showDOI{\tempurl}


\bibitem[\protect\citeauthoryear{Grossberg}{Grossberg}{2013}]%
        {BCF-art-elnabarawy-2016}
\bibfield{author}{\bibinfo{person}{Stephen Grossberg}.}
  \bibinfo{year}{2013}\natexlab{}.
\newblock \showarticletitle{Adaptive Resonance Theory: How a brain learns to
  consciously attend, learn, and recognize a changing world}.
\newblock \bibinfo{journal}{\emph{Neural Networks}}  \bibinfo{volume}{37}
  (\bibinfo{year}{2013}), \bibinfo{pages}{1--47}.
\newblock
\urldef\tempurl%
\url{https://doi.org/10.1016/j.neunet.2012.09.017}
\showDOI{\tempurl}


\bibitem[\protect\citeauthoryear{Gupta and Aggarwal}{Gupta and
  Aggarwal}{2010}]%
        {BC-geneexpression-Gupta-2010}
\bibfield{author}{\bibinfo{person}{Neelima Gupta} {and} \bibinfo{person}{Seema
  Aggarwal}.} \bibinfo{year}{2010}\natexlab{}.
\newblock \showarticletitle{{MIB:} Using mutual information for biclustering
  gene expression data}.
\newblock \bibinfo{journal}{\emph{Pattern Recognit.}} \bibinfo{volume}{43},
  \bibinfo{number}{8} (\bibinfo{year}{2010}), \bibinfo{pages}{2692--2697}.
\newblock
\urldef\tempurl%
\url{https://doi.org/10.1016/j.patcog.2010.03.002}
\showDOI{\tempurl}


\bibitem[\protect\citeauthoryear{Harper and Konstan}{Harper and
  Konstan}{2016}]%
        {Dataset-Movielens-2016}
\bibfield{author}{\bibinfo{person}{F.~Maxwell Harper} {and}
  \bibinfo{person}{Joseph~A. Konstan}.} \bibinfo{year}{2016}\natexlab{}.
\newblock \showarticletitle{The MovieLens Datasets: History and Context}.
\newblock \bibinfo{journal}{\emph{{ACM} Trans. Interact. Intell. Syst.}}
  \bibinfo{volume}{5}, \bibinfo{number}{4} (\bibinfo{year}{2016}),
  \bibinfo{pages}{19:1--19:19}.
\newblock
\urldef\tempurl%
\url{https://doi.org/10.1145/2827872}
\showDOI{\tempurl}


\bibitem[\protect\citeauthoryear{Hartigan}{Hartigan}{1972}]%
        {BC-Hartigan-partitionbiclust-1972}
\bibfield{author}{\bibinfo{person}{J.~A. Hartigan}.}
  \bibinfo{year}{1972}\natexlab{}.
\newblock \showarticletitle{Direct Clustering of a Data Matrix}.
\newblock \bibinfo{journal}{\emph{J. Amer. Statist. Assoc.}}
  \bibinfo{volume}{67}, \bibinfo{number}{337} (\bibinfo{year}{1972}),
  \bibinfo{pages}{123--129}.
\newblock
\urldef\tempurl%
\url{https://doi.org/10.1080/01621459.1972.10481214}
\showDOI{\tempurl}
\showeprint{https://www.tandfonline.com/doi/pdf/10.1080/01621459.1972.10481214}


\bibitem[\protect\citeauthoryear{Henriques, Antunes, and Madeira}{Henriques
  et~al\mbox{.}}{2015}]%
        {mypr}
\bibfield{author}{\bibinfo{person}{Rui Henriques}, \bibinfo{person}{Claudia
  Antunes}, {and} \bibinfo{person}{Sara~C. Madeira}.}
  \bibinfo{year}{2015}\natexlab{}.
\newblock \showarticletitle{A {S}tructured {V}iew on {P}attern {M}ining-based
  {B}iclustering}.
\newblock \bibinfo{journal}{\emph{Pattern Recognition}} \bibinfo{volume}{4},
  \bibinfo{number}{12} (\bibinfo{year}{2015}), \bibinfo{pages}{3941–--3958}.
\newblock


\bibitem[\protect\citeauthoryear{Henriques, Ferreira, and Madeira}{Henriques
  et~al\mbox{.}}{2017}]%
        {BC-BicPams-Rui-2017}
\bibfield{author}{\bibinfo{person}{Rui Henriques},
  \bibinfo{person}{Francisco~L. Ferreira}, {and} \bibinfo{person}{Sara~C.
  Madeira}.} \bibinfo{year}{2017}\natexlab{}.
\newblock \showarticletitle{BicPAMS: software for biological data analysis with
  pattern-based biclustering}.
\newblock \bibinfo{journal}{\emph{{BMC} Bioinform.}} \bibinfo{volume}{18},
  \bibinfo{number}{1} (\bibinfo{year}{2017}), \bibinfo{pages}{82:1--82:16}.
\newblock
\urldef\tempurl%
\url{https://doi.org/10.1186/s12859-017-1493-3}
\showDOI{\tempurl}


\bibitem[\protect\citeauthoryear{Henriques and Madeira}{Henriques and
  Madeira}{2014}]%
        {henriques2014bicpam}
\bibfield{author}{\bibinfo{person}{Rui Henriques} {and} \bibinfo{person}{Sara~C
  Madeira}.} \bibinfo{year}{2014}\natexlab{}.
\newblock \showarticletitle{BicPAM: Pattern-based biclustering for biomedical
  data analysis}.
\newblock \bibinfo{journal}{\emph{Algorithms for Molecular Biology}}
  \bibinfo{volume}{9}, \bibinfo{number}{1} (\bibinfo{year}{2014}),
  \bibinfo{pages}{1--30}.
\newblock


\bibitem[\protect\citeauthoryear{Henriques and Madeira}{Henriques and
  Madeira}{2016}]%
        {BC-BIC2PAM-Rui-2016}
\bibfield{author}{\bibinfo{person}{Rui Henriques} {and}
  \bibinfo{person}{Sara~C. Madeira}.} \bibinfo{year}{2016}\natexlab{}.
\newblock \showarticletitle{BiC2PAM: constraint-guided biclustering for
  biological data analysis with domain knowledge}.
\newblock \bibinfo{journal}{\emph{Algorithms Mol. Biol.}}  \bibinfo{volume}{11}
  (\bibinfo{year}{2016}), \bibinfo{pages}{23}.
\newblock
\urldef\tempurl%
\url{https://doi.org/10.1186/s13015-016-0085-5}
\showDOI{\tempurl}


\bibitem[\protect\citeauthoryear{Henriques and Madeira}{Henriques and
  Madeira}{2018}]%
        {Henriques2017}
\bibfield{author}{\bibinfo{person}{Rui Henriques} {and} \bibinfo{person}{Sara~C
  Madeira}.} \bibinfo{year}{2018}\natexlab{}.
\newblock \showarticletitle{BSig: evaluating the statistical significance of
  biclustering solutions}.
\newblock \bibinfo{journal}{\emph{Data Mining and Knowledge Discovery}}
  \bibinfo{volume}{32}, \bibinfo{number}{1} (\bibinfo{year}{2018}),
  \bibinfo{pages}{124--161}.
\newblock


\bibitem[\protect\citeauthoryear{Herlocker, Konstan, Terveen, and
  Riedl}{Herlocker et~al\mbox{.}}{2004}]%
        {CF-EvaluateCF-Herlocker-2004}
\bibfield{author}{\bibinfo{person}{Jonathan~L. Herlocker},
  \bibinfo{person}{Joseph~A. Konstan}, \bibinfo{person}{Loren~G. Terveen},
  {and} \bibinfo{person}{John Riedl}.} \bibinfo{year}{2004}\natexlab{}.
\newblock \showarticletitle{Evaluating collaborative filtering recommender
  systems}.
\newblock \bibinfo{journal}{\emph{{ACM} Trans. Inf. Syst.}}
  \bibinfo{volume}{22}, \bibinfo{number}{1} (\bibinfo{year}{2004}),
  \bibinfo{pages}{5--53}.
\newblock
\urldef\tempurl%
\url{https://doi.org/10.1145/963770.963772}
\showDOI{\tempurl}


\bibitem[\protect\citeauthoryear{Hug}{Hug}{2020}]%
        {Surpriselib-Hug-2020}
\bibfield{author}{\bibinfo{person}{Nicolas Hug}.}
  \bibinfo{year}{2020}\natexlab{}.
\newblock \showarticletitle{Surprise: {A} Python library for recommender
  systems}.
\newblock \bibinfo{journal}{\emph{J. Open Source Softw.}} \bibinfo{volume}{5},
  \bibinfo{number}{52} (\bibinfo{year}{2020}), \bibinfo{pages}{2174}.
\newblock
\urldef\tempurl%
\url{https://doi.org/10.21105/joss.02174}
\showDOI{\tempurl}


\bibitem[\protect\citeauthoryear{Irish}{Irish}{2010}]%
        {BCF-biclustcffusion-kantmahara-2017-MMD-SIMMEASURE}
\bibfield{author}{\bibinfo{person}{Joel~D Irish}.}
  \bibinfo{year}{2010}\natexlab{}.
\newblock \showarticletitle{The mean measure of divergence: Its utility in
  model-free and model-bound analyses relative to the Mahalanobis D2 distance
  for nonmetric traits}.
\newblock \bibinfo{journal}{\emph{American Journal of Human Biology}}
  \bibinfo{volume}{22}, \bibinfo{number}{3} (\bibinfo{year}{2010}),
  \bibinfo{pages}{378--395}.
\newblock


\bibitem[\protect\citeauthoryear{Juniarta}{Juniarta}{2019}]%
        {BC-biclustneighborhoodtopn-FCA-PHDTHESIS-juniarta2019}
\bibfield{author}{\bibinfo{person}{Nyoman Juniarta}.}
  \bibinfo{year}{2019}\natexlab{}.
\newblock \emph{\bibinfo{title}{Mining complex data and biclustering using
  formal concept analysis. (Fouille de donn{\'{e}}es complexes et biclustering
  avec l'analyse formelle de concepts)}}.
\newblock \bibinfo{thesistype}{Ph.D. Dissertation}. \bibinfo{school}{University
  of Lorraine, Nancy, France}.
\newblock
\urldef\tempurl%
\url{https://tel.archives-ouvertes.fr/tel-02426034}
\showURL{%
\tempurl}


\bibitem[\protect\citeauthoryear{Kant and Mahara}{Kant and Mahara}{2018}]%
        {BCF-biclustcffusion-kantmahara-2017}
\bibfield{author}{\bibinfo{person}{Surya Kant} {and} \bibinfo{person}{Tripti
  Mahara}.} \bibinfo{year}{2018}\natexlab{}.
\newblock \showarticletitle{Nearest biclusters collaborative filtering
  framework with fusion}.
\newblock \bibinfo{journal}{\emph{J. Comput. Sci.}}  \bibinfo{volume}{25}
  (\bibinfo{year}{2018}), \bibinfo{pages}{204--212}.
\newblock
\urldef\tempurl%
\url{https://doi.org/10.1016/j.jocs.2017.03.018}
\showDOI{\tempurl}


\bibitem[\protect\citeauthoryear{Kluger, Basri, Chang, and Gerstein}{Kluger
  et~al\mbox{.}}{2003}]%
        {kluger2003spectral}
\bibfield{author}{\bibinfo{person}{Yuval Kluger}, \bibinfo{person}{Ronen
  Basri}, \bibinfo{person}{Joseph~T Chang}, {and} \bibinfo{person}{Mark
  Gerstein}.} \bibinfo{year}{2003}\natexlab{}.
\newblock \showarticletitle{Spectral biclustering of microarray data:
  coclustering genes and conditions}.
\newblock \bibinfo{journal}{\emph{Genome research}} \bibinfo{volume}{13},
  \bibinfo{number}{4} (\bibinfo{year}{2003}), \bibinfo{pages}{703--716}.
\newblock


\bibitem[\protect\citeauthoryear{Koren}{Koren}{2008}]%
        {MatrixFactCF-svd++-koren2008}
\bibfield{author}{\bibinfo{person}{Yehuda Koren}.}
  \bibinfo{year}{2008}\natexlab{}.
\newblock \showarticletitle{Factorization meets the neighborhood: a
  multifaceted collaborative filtering model}. In
  \bibinfo{booktitle}{\emph{Proceedings of the 14th {ACM} {SIGKDD}
  International Conference on Knowledge Discovery and Data Mining, Las Vegas,
  Nevada, USA, August 24-27, 2008}}, \bibfield{editor}{\bibinfo{person}{Ying
  Li}, \bibinfo{person}{Bing Liu}, {and} \bibinfo{person}{Sunita Sarawagi}}
  (Eds.). \bibinfo{publisher}{{ACM}}, \bibinfo{pages}{426--434}.
\newblock
\urldef\tempurl%
\url{https://doi.org/10.1145/1401890.1401944}
\showDOI{\tempurl}


\bibitem[\protect\citeauthoryear{Koren}{Koren}{2010}]%
        {BaselineCF-bias-Koren2010}
\bibfield{author}{\bibinfo{person}{Yehuda Koren}.}
  \bibinfo{year}{2010}\natexlab{}.
\newblock \showarticletitle{Factor in the neighbors: Scalable and accurate
  collaborative filtering}.
\newblock \bibinfo{journal}{\emph{{ACM} Trans. Knowl. Discov. Data}}
  \bibinfo{volume}{4}, \bibinfo{number}{1} (\bibinfo{year}{2010}),
  \bibinfo{pages}{1:1--1:24}.
\newblock
\urldef\tempurl%
\url{https://doi.org/10.1145/1644873.1644874}
\showDOI{\tempurl}


\bibitem[\protect\citeauthoryear{Li, Ma, Tang, Paterson, and Xu}{Li
  et~al\mbox{.}}{2009}]%
        {BC-QUIBIC-2009}
\bibfield{author}{\bibinfo{person}{Guojun Li}, \bibinfo{person}{Qin Ma},
  \bibinfo{person}{Haibao Tang}, \bibinfo{person}{Andrew~H. Paterson}, {and}
  \bibinfo{person}{Ying Xu}.} \bibinfo{year}{2009}\natexlab{}.
\newblock \showarticletitle{{QUBIC: a qualitative biclustering algorithm for
  analyses of gene expression data}}.
\newblock \bibinfo{journal}{\emph{Nucleic Acids Research}}
  \bibinfo{volume}{37}, \bibinfo{number}{15} (\bibinfo{date}{06}
  \bibinfo{year}{2009}), \bibinfo{pages}{e101--e101}.
\newblock
\showISSN{0305-1048}
\urldef\tempurl%
\url{https://doi.org/10.1093/nar/gkp491}
\showDOI{\tempurl}


\bibitem[\protect\citeauthoryear{Liang and Leng}{Liang and Leng}{2014}]%
        {BCF-LiangLeng-2014}
\bibfield{author}{\bibinfo{person}{Changyong Liang} {and}
  \bibinfo{person}{Yajun Leng}.} \bibinfo{year}{2014}\natexlab{}.
\newblock \showarticletitle{Collaborative filtering based on
  information-theoretic co-clustering}.
\newblock \bibinfo{journal}{\emph{Int. J. Syst. Sci.}} \bibinfo{volume}{45},
  \bibinfo{number}{3} (\bibinfo{year}{2014}), \bibinfo{pages}{589--597}.
\newblock
\urldef\tempurl%
\url{https://doi.org/10.1080/00207721.2012.724109}
\showDOI{\tempurl}


\bibitem[\protect\citeauthoryear{Luo, Zhou, Xia, and Zhu}{Luo
  et~al\mbox{.}}{2014}]%
        {MatrixFactCF-NMF-Xin-2014}
\bibfield{author}{\bibinfo{person}{Xin Luo}, \bibinfo{person}{Mengchu Zhou},
  \bibinfo{person}{Yunni Xia}, {and} \bibinfo{person}{Qingsheng Zhu}.}
  \bibinfo{year}{2014}\natexlab{}.
\newblock \showarticletitle{An Efficient Non-Negative
  Matrix-Factorization-Based Approach to Collaborative Filtering for
  Recommender Systems}.
\newblock \bibinfo{journal}{\emph{{IEEE} Trans. Ind. Informatics}}
  \bibinfo{volume}{10}, \bibinfo{number}{2} (\bibinfo{year}{2014}),
  \bibinfo{pages}{1273--1284}.
\newblock
\urldef\tempurl%
\url{https://doi.org/10.1109/TII.2014.2308433}
\showDOI{\tempurl}


\bibitem[\protect\citeauthoryear{Madeira and Oliveira}{Madeira and
  Oliveira}{2004}]%
        {BC-Survey-Sara-2004}
\bibfield{author}{\bibinfo{person}{Sara~C. Madeira} {and}
  \bibinfo{person}{Arlindo~L. Oliveira}.} \bibinfo{year}{2004}\natexlab{}.
\newblock \showarticletitle{Biclustering Algorithms for Biological Data
  Analysis: {A} Survey}.
\newblock \bibinfo{journal}{\emph{{IEEE} {ACM} Trans. Comput. Biol.
  Bioinform.}} \bibinfo{volume}{1}, \bibinfo{number}{1} (\bibinfo{year}{2004}),
  \bibinfo{pages}{24--45}.
\newblock
\urldef\tempurl%
\url{https://doi.org/10.1109/TCBB.2004.2}
\showDOI{\tempurl}


\bibitem[\protect\citeauthoryear{Mahara et~al\mbox{.}}{Mahara
  et~al\mbox{.}}{2016}]%
        {BCF-biclustcffusion-kantmahara-2017-CjacMD-SIMMEASURE}
\bibfield{author}{\bibinfo{person}{Tripti Mahara} {et~al\mbox{.}}}
  \bibinfo{year}{2016}\natexlab{}.
\newblock \showarticletitle{A new similarity measure based on mean measure of
  divergence for collaborative filtering in sparse environment}.
\newblock \bibinfo{journal}{\emph{Procedia Computer Science}}
  \bibinfo{volume}{89} (\bibinfo{year}{2016}), \bibinfo{pages}{450--456}.
\newblock


\bibitem[\protect\citeauthoryear{Murali and Kasif}{Murali and Kasif}{2003}]%
        {BC-xMotifs-Murali-2003}
\bibfield{author}{\bibinfo{person}{T.~M. Murali} {and} \bibinfo{person}{Simon
  Kasif}.} \bibinfo{year}{2003}\natexlab{}.
\newblock \showarticletitle{Extracting Conserved Gene Expression Motifs from
  Gene Expression Data}. In \bibinfo{booktitle}{\emph{Proceedings of the 8th
  Pacific Symposium on Biocomputing, {PSB} 2003, Lihue, Hawaii, USA, January
  3-7, 2003}}, \bibfield{editor}{\bibinfo{person}{Russ~B. Altman},
  \bibinfo{person}{A.~Keith Dunker}, \bibinfo{person}{Lawrence Hunter}, {and}
  \bibinfo{person}{Teri~E. Klein}} (Eds.). \bibinfo{pages}{77--88}.
\newblock
\urldef\tempurl%
\url{http://psb.stanford.edu/psb-online/proceedings/psb03/murali.pdf}
\showURL{%
\tempurl}


\bibitem[\protect\citeauthoryear{Resnick, Iacovou, Suchak, Bergstrom, and
  Riedl}{Resnick et~al\mbox{.}}{1994}]%
        {UserBasedCF-Resnick-1994}
\bibfield{author}{\bibinfo{person}{Paul Resnick}, \bibinfo{person}{Neophytos
  Iacovou}, \bibinfo{person}{Mitesh Suchak}, \bibinfo{person}{Peter Bergstrom},
  {and} \bibinfo{person}{John Riedl}.} \bibinfo{year}{1994}\natexlab{}.
\newblock \showarticletitle{GroupLens: An Open Architecture for Collaborative
  Filtering of Netnews}. In \bibinfo{booktitle}{\emph{{CSCW} '94, Proceedings
  of the Conference on Computer Supported Cooperative Work, Chapel Hill, NC,
  USA, October 22-26, 1994}}, \bibfield{editor}{\bibinfo{person}{John~B.
  Smith}, \bibinfo{person}{F.~Donelson Smith}, {and} \bibinfo{person}{Thomas~W.
  Malone}} (Eds.). \bibinfo{publisher}{{ACM}}, \bibinfo{pages}{175--186}.
\newblock
\urldef\tempurl%
\url{https://doi.org/10.1145/192844.192905}
\showDOI{\tempurl}


\bibitem[\protect\citeauthoryear{Ricci, Rokach, and Shapira}{Ricci
  et~al\mbox{.}}{2015}]%
        {CF-RecommenderSystemshandbook-Ricci-2015}
\bibfield{editor}{\bibinfo{person}{Francesco Ricci}, \bibinfo{person}{Lior
  Rokach}, {and} \bibinfo{person}{Bracha Shapira}} (Eds.).
  \bibinfo{year}{2015}\natexlab{}.
\newblock \bibinfo{booktitle}{\emph{Recommender Systems Handbook}}.
\newblock \bibinfo{publisher}{Springer}.
\newblock
\showISBNx{978-1-4899-7636-9}
\urldef\tempurl%
\url{https://doi.org/10.1007/978-1-4899-7637-6}
\showDOI{\tempurl}


\bibitem[\protect\citeauthoryear{Sarwar, Karypis, Konstan, and Riedl}{Sarwar
  et~al\mbox{.}}{2002}]%
        {ClusteringCF-Sarwar-2002}
\bibfield{author}{\bibinfo{person}{Badrul~M Sarwar}, \bibinfo{person}{George
  Karypis}, \bibinfo{person}{Joseph Konstan}, {and} \bibinfo{person}{John
  Riedl}.} \bibinfo{year}{2002}\natexlab{}.
\newblock \showarticletitle{Recommender systems for large-scale e-commerce:
  Scalable neighborhood formation using clustering}. In
  \bibinfo{booktitle}{\emph{Proceedings of the fifth international conference
  on computer and information technology}}, Vol.~\bibinfo{volume}{1}. Citeseer,
  \bibinfo{pages}{291--324}.
\newblock


\bibitem[\protect\citeauthoryear{Sarwar, Karypis, Konstan, and Riedl}{Sarwar
  et~al\mbox{.}}{2001}]%
        {ItemBasedCF-Sarwar-2001}
\bibfield{author}{\bibinfo{person}{Badrul~Munir Sarwar},
  \bibinfo{person}{George Karypis}, \bibinfo{person}{Joseph~A. Konstan}, {and}
  \bibinfo{person}{John Riedl}.} \bibinfo{year}{2001}\natexlab{}.
\newblock \showarticletitle{Item-based collaborative filtering recommendation
  algorithms}. In \bibinfo{booktitle}{\emph{Proceedings of the Tenth
  International World Wide Web Conference, {WWW} 10, Hong Kong, China, May 1-5,
  2001}}, \bibfield{editor}{\bibinfo{person}{Vincent~Y. Shen},
  \bibinfo{person}{Nobuo Saito}, \bibinfo{person}{Michael~R. Lyu}, {and}
  \bibinfo{person}{Mary~Ellen Zurko}} (Eds.). \bibinfo{publisher}{{ACM}},
  \bibinfo{pages}{285--295}.
\newblock
\urldef\tempurl%
\url{https://doi.org/10.1145/371920.372071}
\showDOI{\tempurl}


\bibitem[\protect\citeauthoryear{Sim, Gopalkrishnan, Zimek, and Cong}{Sim
  et~al\mbox{.}}{2013}]%
        {BC-Survey-Kelvin-2013}
\bibfield{author}{\bibinfo{person}{Kelvin Sim}, \bibinfo{person}{Vivekanand
  Gopalkrishnan}, \bibinfo{person}{Arthur Zimek}, {and} \bibinfo{person}{Gao
  Cong}.} \bibinfo{year}{2013}\natexlab{}.
\newblock \showarticletitle{A survey on enhanced subspace clustering}.
\newblock \bibinfo{journal}{\emph{Data Min. Knowl. Discov.}}
  \bibinfo{volume}{26}, \bibinfo{number}{2} (\bibinfo{year}{2013}),
  \bibinfo{pages}{332--397}.
\newblock
\urldef\tempurl%
\url{https://doi.org/10.1007/s10618-012-0258-x}
\showDOI{\tempurl}


\bibitem[\protect\citeauthoryear{Singh and Mehrotra}{Singh and
  Mehrotra}{2018}]%
        {BCF-impactbiclusteringcf-Singh-2018}
\bibfield{author}{\bibinfo{person}{Monika Singh} {and} \bibinfo{person}{Monica
  Mehrotra}.} \bibinfo{year}{2018}\natexlab{}.
\newblock \showarticletitle{Impact of biclustering on the performance of
  Biclustering based Collaborative Filtering}.
\newblock \bibinfo{journal}{\emph{Expert Syst. Appl.}}  \bibinfo{volume}{113}
  (\bibinfo{year}{2018}), \bibinfo{pages}{443--456}.
\newblock
\urldef\tempurl%
\url{https://doi.org/10.1016/j.eswa.2018.06.001}
\showDOI{\tempurl}


\bibitem[\protect\citeauthoryear{Su and Khoshgoftaar}{Su and
  Khoshgoftaar}{2009}]%
        {Survey-CF-Su-2009}
\bibfield{author}{\bibinfo{person}{Xiaoyuan Su} {and} \bibinfo{person}{Taghi~M.
  Khoshgoftaar}.} \bibinfo{year}{2009}\natexlab{}.
\newblock \showarticletitle{A Survey of Collaborative Filtering Techniques}.
\newblock \bibinfo{journal}{\emph{Adv. Artif. Intell.}}  \bibinfo{volume}{2009}
  (\bibinfo{year}{2009}), \bibinfo{pages}{421425:1--421425:19}.
\newblock
\urldef\tempurl%
\url{https://doi.org/10.1155/2009/421425}
\showDOI{\tempurl}


\bibitem[\protect\citeauthoryear{Symeonidis, Nanopoulos, Papadopoulos, and
  Manolopoulos}{Symeonidis et~al\mbox{.}}{2006}]%
        {BCF-nearestbicsconst-Symeonidis-2006}
\bibfield{author}{\bibinfo{person}{Panagiotis Symeonidis},
  \bibinfo{person}{Alexandros Nanopoulos}, \bibinfo{person}{Apostolos
  Papadopoulos}, {and} \bibinfo{person}{Yannis Manolopoulos}.}
  \bibinfo{year}{2006}\natexlab{}.
\newblock \showarticletitle{Nearest-Biclusters Collaborative Filtering with
  Constant Values}. In \bibinfo{booktitle}{\emph{Advances in Web Mining and Web
  Usage Analysis, 8th International Workshop on Knowledge Discovery on the Web,
  WebKDD 2006, Philadelphia, PA, USA, August 20, 2006, Revised Papers}}
  \emph{(\bibinfo{series}{Lecture Notes in Computer Science},
  Vol.~\bibinfo{volume}{4811})}, \bibfield{editor}{\bibinfo{person}{Olfa
  Nasraoui}, \bibinfo{person}{Myra Spiliopoulou}, \bibinfo{person}{Jaideep
  Srivastava}, \bibinfo{person}{Bamshad Mobasher}, {and}
  \bibinfo{person}{Brij~M. Masand}} (Eds.). \bibinfo{publisher}{Springer},
  \bibinfo{pages}{36--55}.
\newblock
\urldef\tempurl%
\url{https://doi.org/10.1007/978-3-540-77485-3\_3}
\showDOI{\tempurl}


\bibitem[\protect\citeauthoryear{Symeonidis, Nanopoulos, Papadopoulos, and
  Manolopoulos}{Symeonidis et~al\mbox{.}}{2008}]%
        {BCF-nearestbicsconstcoherent-Symeonidis-2008}
\bibfield{author}{\bibinfo{person}{Panagiotis Symeonidis},
  \bibinfo{person}{Alexandros Nanopoulos}, \bibinfo{person}{Apostolos~N.
  Papadopoulos}, {and} \bibinfo{person}{Yannis Manolopoulos}.}
  \bibinfo{year}{2008}\natexlab{}.
\newblock \showarticletitle{Nearest-biclusters collaborative filtering based on
  constant and coherent values}.
\newblock \bibinfo{journal}{\emph{Inf. Retr.}} \bibinfo{volume}{11},
  \bibinfo{number}{1} (\bibinfo{year}{2008}), \bibinfo{pages}{51--75}.
\newblock
\urldef\tempurl%
\url{https://doi.org/10.1007/s10791-007-9038-4}
\showDOI{\tempurl}


\bibitem[\protect\citeauthoryear{Vinagre, Jorge, and Gama}{Vinagre
  et~al\mbox{.}}{2015}]%
        {CF-ExploitTimeCF-VinagreJoaoJorge-2015}
\bibfield{author}{\bibinfo{person}{Jo{\~{a}}o Vinagre},
  \bibinfo{person}{Al{\'{\i}}pio~M{\'{a}}rio Jorge}, {and}
  \bibinfo{person}{Jo{\~{a}}o Gama}.} \bibinfo{year}{2015}\natexlab{}.
\newblock \showarticletitle{An overview on the exploitation of time in
  collaborative filtering}.
\newblock \bibinfo{journal}{\emph{Wiley Interdiscip. Rev. Data Min. Knowl.
  Discov.}} \bibinfo{volume}{5}, \bibinfo{number}{5} (\bibinfo{year}{2015}),
  \bibinfo{pages}{195--215}.
\newblock
\urldef\tempurl%
\url{https://doi.org/10.1002/widm.1160}
\showDOI{\tempurl}


\bibitem[\protect\citeauthoryear{Xie, Ma, Zhang, Liu, Cao, Wang, Xu, Zhang, and
  Ma}{Xie et~al\mbox{.}}{2020}]%
        {BC-QUBIC2-2019}
\bibfield{author}{\bibinfo{person}{Juan Xie}, \bibinfo{person}{Anjun Ma},
  \bibinfo{person}{Yu Zhang}, \bibinfo{person}{Bingqiang Liu},
  \bibinfo{person}{Sha Cao}, \bibinfo{person}{Cankun Wang},
  \bibinfo{person}{Jennifer Xu}, \bibinfo{person}{Chi Zhang}, {and}
  \bibinfo{person}{Qin Ma}.} \bibinfo{year}{2020}\natexlab{}.
\newblock \showarticletitle{{QUBIC2:} a novel and robust biclustering algorithm
  for analyses and interpretation of large-scale RNA-Seq data}.
\newblock \bibinfo{journal}{\emph{Bioinform.}} \bibinfo{volume}{36},
  \bibinfo{number}{4} (\bibinfo{year}{2020}), \bibinfo{pages}{1143--1149}.
\newblock
\urldef\tempurl%
\url{https://doi.org/10.1093/bioinformatics/btz692}
\showDOI{\tempurl}


\bibitem[\protect\citeauthoryear{Xu and II}{Xu and II}{2011}]%
        {BC-BARTMAP-XU-2011}
\bibfield{author}{\bibinfo{person}{Rui Xu} {and} \bibinfo{person}{Donald
  C.~Wunsch II}.} \bibinfo{year}{2011}\natexlab{}.
\newblock \showarticletitle{{BARTMAP:} {A} viable structure for biclustering}.
\newblock \bibinfo{journal}{\emph{Neural Networks}} \bibinfo{volume}{24},
  \bibinfo{number}{7} (\bibinfo{year}{2011}), \bibinfo{pages}{709--716}.
\newblock
\urldef\tempurl%
\url{https://doi.org/10.1016/j.neunet.2011.03.020}
\showDOI{\tempurl}


\bibitem[\protect\citeauthoryear{Xue, Lin, Yang, Xi, Zeng, Yu, and Chen}{Xue
  et~al\mbox{.}}{2005}]%
        {ClustCF-xue-2005}
\bibfield{author}{\bibinfo{person}{Gui{-}Rong Xue}, \bibinfo{person}{Chenxi
  Lin}, \bibinfo{person}{Qiang Yang}, \bibinfo{person}{Wensi Xi},
  \bibinfo{person}{Hua{-}Jun Zeng}, \bibinfo{person}{Yong Yu}, {and}
  \bibinfo{person}{Zheng Chen}.} \bibinfo{year}{2005}\natexlab{}.
\newblock \showarticletitle{Scalable collaborative filtering using
  cluster-based smoothing}. In \bibinfo{booktitle}{\emph{{SIGIR} 2005:
  Proceedings of the 28th Annual International {ACM} {SIGIR} Conference on
  Research and Development in Information Retrieval, Salvador, Brazil, August
  15-19, 2005}}, \bibfield{editor}{\bibinfo{person}{Ricardo~A. Baeza{-}Yates},
  \bibinfo{person}{Nivio Ziviani}, \bibinfo{person}{Gary Marchionini},
  \bibinfo{person}{Alistair Moffat}, {and} \bibinfo{person}{John Tait}} (Eds.).
  \bibinfo{publisher}{{ACM}}, \bibinfo{pages}{114--121}.
\newblock
\urldef\tempurl%
\url{https://doi.org/10.1145/1076034.1076056}
\showDOI{\tempurl}


\end{thebibliography}

\end{document}